\documentclass[reprint,
 amsmath,amssymb,
 aps,
prstab
]{revtex4-1}

\usepackage{graphicx}
\usepackage{dcolumn}
\usepackage{bm}

\usepackage{amsmath}
\DeclareMathOperator{\sech}{sech}

\usepackage{textcomp} 

\graphicspath{ {fig/} }

\makeatletter
\newcommand\thefontsize[1]{{#1 The current font size is: \f@size pt\par}}
\newcommand\thefontsizeHere{{The current font size is: \f@size pt\par}}
\makeatother

\begin{document}

\preprint{APS/123-QED}

\title{Prediction of Beam Losses during Crab Cavity Quenches at the HL-LHC}


\author{R. Apsimon}
\email{r.apsimon@lancaster.ac.uk}
\author{G. Burt}
\author{A. Dexter}
\author{N. Shipman}
\affiliation{Engineering Department, Lancaster University, LA1 4YW, UK }
\affiliation{Cockcroft Institute, Daresbury Laboratory, Warrington, WA4 4AD, UK}

\author{P. Baudrenghien}
\author{R. Calaga}
\author{A. Castilla}
\author{A. Macpherson}
\author{K. Ness Sjobak}
\author{A. Santamaria Garcia}
\author{N. Stapley}
\affiliation{CERN, Geneva, Switzerland}

\author{A. Alekou}
\email{also at CERN, Geneva, Switzerland}
\author{R. B. Appleby}
\affiliation{School of Physics and Astronomy, The University of Manchester, M13
9PL, UK}
\affiliation{Cockcroft Institute, Daresbury Laboratory, Warrington, WA4 4AD, UK}

\date{\today}

\begin{abstract}
Studies of the crab cavities at KEKB revealed that the RF phase could shift by up to 50$^{\circ}$ within $\sim$50~$\mu$s during a quench; while the cavity voltage is still at approximately 75\% of its nominal amplitude. If such a failure were to occur on the HL-LHC crab cavities, it is likely that the machine would sustain substantial damage to the beam line and surrounding infrastructure due to uncontrolled beam loss before the machine protection system could dump the beam. We have developed a low-level RF system model, including detuning mechanisms and beam loading, and use this to simulate the behaviour of a crab cavity during a quench, modeling the low-level RF system, detuning mechanisms and beam loading. We supplement this with measurement data of the actual RF response of the proof of principle Double-Quarter Wave Crab Cravity during a quench. Extrapolating these measurements to the HL-LHC, we show that Lorentz Force detuning is the dominant effect leading to phase shifts in the crab cavity during quenches; rather than pressure detuning which is expected to be dominant for the KEKB crab cavities. The total frequency shift for the HL-LHC crab cavities during quenches is expected to be about 460~Hz, leading to a phase shift of no more than 3$^{\circ}$. The results of the quench model are read into a particle tracking simulation, SixTrack, and used to determine the effect of quenches on the HL-LHC beam. The quench model has been benchmarked against the KEKB experimental measurements. In this paper we present the results of the simulations on a crab cavity failure for HL-LHC as well as for the SPS and show that beam loss is negligible when using a realistic low-level RF response.
\end{abstract}

\maketitle

\section{Introduction}
\label{sec:introduction}

The High Luminosity upgrade for the LHC (HL-LHC)~\cite{Apollinari:2116337} will use crab cavities to compensate for the luminosity reduction at the interaction points (IPs) due to the crossing angle of the counter-rotating bunch trains. A crab cavity is a deflecting cavity, phased such that the centroid of the beam passes through the cavity on the zero-crossing; hence providing an effective rotation to the bunches. If the phase of the cavity field changes with respect to the bunches, then the bunch centroid experiences a kick, which can lead to particle losses. During a quench, the phase and amplitude of the cavity field can change relatively quickly, potentially leading to beam losses throughout the ring.

The response time of the machine protection system depends on many different factors, but has an upper limit of 3 turns of the LHC ring, or approximately 267~\textmu{s}~\cite{LHCdump}. Beam losses which occur within 267~\textmu{s} of a failure can potentially damage the LHC beam line.

Measurements of the crab cavities installed at KEKB have revealed that the RF phase shifts by up to 50$^{\circ}$ within 50~\textmu{s} during a quench; while the cavity voltage is still approximately 75\% of its nominal value~\cite{KEKBcrab}. Should such a failure occur on the HL-LHC crab cavities, the beam losses could cause substantial damage to the beam line. The following studies are critical as CERN could not risk installing crab cavities in HL-LHC if they have the potential to damage the machine; which would severely limit the luminosity upgrade for HL-LHC.

Several design studies for the HL-LHC crab cavities include the double-quarter (DQW) wave~\cite{PhysRevSTAB.18.041004}, RF Dipole (RFD)~\cite{PhysRevSTAB.16.082001, PhysRevSTAB.16.012004} and the four-rod crab cavity (4RCC)~\cite{hallPRAB}; although the 4RCC was not selected to be one of the final designs. Before installing the crab cavities in HL-LHC, prototypes of the DQW and RFD crab cavities will first be installed in SPS for testing. Therefore we simulate the impact on the beam due to a crab cavity quench in the SPS in order to compare to measurements during SPS tests. The comparison of simulation and measurement will be used to verify the HL-LHC simulations ahead of the installation of the HL-LHC crab cavities.

Previous studies have imposed an arbitrary phase and amplitude response on the crab cavities during a quench to study beam losses~\cite{Andrea15}. In this paper we utilise a more realistic RF system model, including low-level RF (LLRF), pressure, Lorentz Force and resistive detuning mechanisms, as well as microphonics and beam loading in order to obtain a more realistic phase and amplitude evolution during crab cavity quenches.

\section{\label{sec:cavmodel}Modeling the RF system}
\subsection{Cavity model}
We shall consider an RF cavity driven by an RF source; which we shall assume to be a tetrode. We shall assume that the tetrode excites a total of $N$ modes in the cavity. In this case, the cavity voltage for the $i^{\text{th}}$ mode is governed by the second order ODE given in Eq.~\ref{eq:caveq1}~\cite{dexter}.

\begin{equation}
\frac{\mathrm{d}^{2}\mathbf{V}_{i}}{\mathrm{d}t^{2}}+\frac{\omega_{i}}{Q_{0,i}}\frac{\mathrm{d}\mathbf{V}_{i}}{\mathrm{d}t}+\frac{\omega_{i}}{Q_{e,i}}\sum_{k=1}^{N}{\frac{\mathrm{d}\mathbf{V}_{k}}{\mathrm{d}t}}+\omega_{i}^{2}\mathbf{V}_{i}=\frac{2\omega_{i}}{Q_{e,i}}\frac{\mathrm{d}}{\mathrm{d}t}\left[\mathbf{F}e^{-j\omega{t}}\right]
	\label{eq:caveq1}
\end{equation}

Where $Q_{0,i}$ and $Q_{e,i}$ are the intrinsic and external quality factors for the $i^{\text{th}}$~mode respectively, $k$ is the summation index over all modes and $\omega_{i}$ is the resonant angular frequency of the $i^{\text{th}}$~mode. The bold font is used to indicate that the functions are complex numbers and therefore vectors. $\mathbf{F}$ is the voltage corresponding to the input power, $\mathbf{P}_{in}$ from the tetrode, expressed as

\begin{equation}
\mathbf{F}=\sqrt{2\left(\frac{R}{Q}\right)Q_{L}\mathbf{P}_{in}}
	\label{eq:caveq7}.
\end{equation}

Where $Q_{L}$ is the loaded quality factor, given as $\frac{1}{Q_{L}}=\frac{1}{Q_{0}}+\frac{1}{Q_{e}}$ and $\left(R/Q\right)$ is the shunt impedance divided by intrinsic Q-factor. For most cases, we can assume that only one mode is excited in the cavity. This is because the mode separation and bandwidth of the nearby modes are far away from the frequencies excited by the amplifier~\cite{apsimon2017improved,hallPRAB}. In this case, Eq.~\ref{eq:caveq1} is simplified to


\begin{equation}
\frac{\mathrm{d}^{2}\mathbf{V}}{\mathrm{d}t^{2}}+\frac{\omega_{0}}{Q_{L}}\frac{\mathrm{d}\mathbf{V}}{\mathrm{d}t}+\omega_{0}^{2}\mathbf{V}=\frac{2\omega_{0}}{Q_{e}}\frac{\mathrm{d}}{\mathrm{d}t}\left[\mathbf{F}e^{-j\omega{t}}\right]
	\label{eq:caveq2}
\end{equation}

\noindent where the subscript $i$ has been removed as we only consider one mode and $\omega_{i}$ replaced with $\omega_{0}$ to be consistent with conventional notation.

Under steady state conditions, Eq.~\ref{eq:caveq2} is the well known equation of motion for the driven, damped harmonic oscillator, however, during a quench or other type of failure, this is not true. During a quench, $Q_{0}$ hence $Q_{L}$ decreases due to increased resistive losses. In addition, detuning mechanisms, such as Lorentz detuning~\cite{LorentzDetuning}, will change $\omega_{0}$ over time. Eq.~\ref{eq:caveq2} can be solved numerically if we have a model for $Q_{0}\left(t\right)$ and $\omega_{0}\left(t\right)$ during a quench.

Although Eq.~\ref{eq:caveq2} can be solved numerically, to study the performance of the LLRF system, the RF phase must be calculated with accuracy on the order of millidegrees as this is the expected stability of the LLRF system. To achieve this, a low-order numerical integrator, such as a 4$^{\text{th}}$-order Runge-Kutta method (RK4), would need to evaluate the cavity voltage hundreds or thousands of times per RF cycle. Alternatively, a higher order integrator, such as RK12, would require evaluating the cavity voltage less often, but each evaluation step would require more computations. In either case, solving Eq.~\ref{eq:caveq2} accurately over hundreds or thousands of RF cycles requires substantial computing time.

In order to overcome the challenges of evaluating Eq.~\ref{eq:caveq2} accurately over many RF cycles, we can make some approximations to derive an envelope equation. This allows us to calculate the cavity voltage once every RF cycle, while still providing the same degree of accuracy. We shall assume that the solution to Eq.~\ref{eq:caveq2} can be expressed as

\begin{equation}
\mathbf{V}\left(t\right)=\mathbf{V}_{\mathrm{cav}}\left(t\right)e^{-j\omega{t}}
	\label{eq:caveq3},
\end{equation}

\noindent relative to a reference frequency $\omega$, where $\mathbf{V}_{\mathrm{cav}}\left(t\right)$ is the complex envelope as a function of time, which gives information of the amplitude and phase of $\mathbf{V}\left(t\right)$ and is assumed to be a slowly varying complex function. By differentiating Eq.~\ref{eq:caveq3}, we obtain

\begin{equation}
\begin{array}{l}
\frac{\mathrm{d}\mathbf{V}}{\mathrm{d}t} = \left[\dot{\mathbf{V}}_{\mathrm{cav}}-j\omega{\mathbf{V}_{\mathrm{cav}}}\right]e^{-j\omega{t}} \\ \\
\frac{\mathrm{d}^{2}\mathbf{V}}{\mathrm{d}t^{2}} = \left[\ddot{\mathbf{V}}_{\mathrm{cav}}-2j\omega\dot{\mathbf{V}}_{\mathrm{cav}}-\omega^{2}\mathbf{V}_{\mathrm{cav}}\right]e^{-j\omega{t}}
\end{array}
\label{eq:caveq4}.
\end{equation}

By substituting Eq.~\ref{eq:caveq4} into Eq.~\ref{eq:caveq2} and canceling the $e^{-j\omega{t}}$ term throughout, we obtain

\begin{equation}
\begin{array}{l}
\ddot{\mathbf{V}}_{\mathrm{cav}}+\left(\frac{\omega_{0}}{Q_{L}}-2j\omega\right)\dot{\mathbf{V}}_{\mathrm{cav}}+\left[\left(\omega_{0}^{2}-\omega^{2}\right)-j\frac{\omega\omega_{0}}{Q_{L}}\right]\mathbf{V}_{\mathrm{cav}} \\ \\
= \frac{2\omega_{0}}{Q_{e}}\left[\dot{\mathbf{F}}-j\omega\mathbf{F}\right]
\end{array}
	\label{eq:caveq5}.
\end{equation}

As we assume that $\mathbf{V}_{\mathrm{cav}}$ is slowly varying, we can neglect the $\ddot{\mathbf{V}}_{\mathrm{cav}}$ term. For almost all RF cavities, $Q_{L}\gg1$, hence by multiplying Eq.~\ref{eq:caveq5} by $\left(\frac{\omega_{0}}{Q_{L}}+2j\omega\right)$ and neglecting terms dependent on $Q_{L}^{-2}$, we obtain

\begin{equation}
\frac{\dot{\mathbf{V}}_{\mathrm{cav}}}{\omega_{0}}+\left[\frac{\omega_{0}^{2}+\omega^{2}}{4Q_{L}\omega^{2}}+j\frac{\omega_{0}^{2}-\omega^{2}}{2\omega\omega_{0}}\right]\mathbf{V}_{\mathrm{cav}} = \frac{j\dot{\mathbf{F}}+\omega\mathbf{F}}{\omega{Q_{e}}}
	\label{eq:caveq6}.
\end{equation}

It should be noted that Eq.~\ref{eq:caveq6} is valid for any mode, hence it can be applied for accelerating cavities as well as deflecting or crabbing cavities.

\subsection{Frequency detuning mechanisms}
In order to determine the cavity voltage from Eq.~\ref{eq:caveq6}, we first need to understand how to model the coefficients and RHS (driving terms). $\omega_{0}$ and $Q_{e}$ are constants, $\omega$ is affected by the cavity detuning mechanism, $Q_{L}$ depends on the behaviour of the quench and the material properties and $\mathbf{F}$ on the LLRF system.

In a superconducting RF (SRF), from Eq.~\ref{eq:caveq6}, the high loaded Q-factor ($\gtrsim10^{5}$) suppresses the real part of the complex coefficient on the left hand side of the equation. Hence a small frequency change can result in a relatively large phase shift. For normal conducting RF, the lower Q-factors result in less sensitivity to frequency offsets, thus detuning is often neglected. For the simulation model, we consider four detuning mechanisms; namely resistive, Lorentz, pressure and microphonics.

\subsubsection{Resistive detuning}
If we consider the undriven form of Eq.~\ref{eq:caveq2}, we have

\begin{equation}
\frac{\mathrm{d}^{2}\mathbf{V}}{\mathrm{d}t^{2}}+\frac{\omega_{0}}{Q_{L}}\frac{\mathrm{d}\mathbf{V}}{\mathrm{d}t}+\omega_{0}^{2}\mathbf{V}=0
	\label{eq:reseq1}.
\end{equation}

The general solution to this is

\begin{equation}
\mathbf{V}=\left[A\cos{\omega{t}}+B\sin{\omega{t}}\right]e^{-\frac{\omega_{0}t}{2Q_{L}}}
	\label{eq:reseq2}
\end{equation}

\noindent where

\begin{equation}
\omega=\omega_{0}\sqrt{1-\frac{1}{4Q_{L}^{2}}}
	\label{eq:reseq3}.
\end{equation}

Therefore the resistive frequency shift is

\begin{equation}
\delta{f}_{R}=f-f_{0}=f_{0}\left(\sqrt{1-\frac{1}{4Q_{L}^{2}}}-1\right)
	\label{eq:reseq4}.
\end{equation}

\subsubsection{Lorentz force detuning}

The electromagnetic fields in an RF cavity apply a radiation pressure to the surface of the cavity given as

\begin{equation}
P_{L} = \frac{\mu_{0}H^{2}-\varepsilon_{0}E^{2}}{4}
	\label{eq:loreq0}
\end{equation}

\noindent due to the Lorentz force~\cite{LorentzDetuning}. The stored energy in the cavity scales with $E^{2}$, $H^{2}$ and $\left|\mathbf{V}\right|^{2}$. The radiation pressure deforms the cavity and therefore changes the resonant frequency. The frequency shift is linearly proportional to the radiation pressure, therefore the Lorentz force detuning can be expressed as

\begin{equation}
\delta{f}_{L}=-K_{L}\left(\left|\mathbf{V}\right|^{2}-\left|\mathbf{V}_{\text{nominal}}\right|^{2}\right)
	\label{eq:loreq1},
\end{equation}

\noindent where $K_{L}$ is the Lorentz force detuning coefficient. Typically, SRF structures are designed such that they operate at the correct frequency when they are under normal operating conditions, therefore the frequency shift is taken as the change in cavity voltage from its nominal value, $\mathbf{V}_{\text{nominal}}$. As the Lorentz force detuning depends on the cavity voltage, this implies that Eq.~\ref{eq:caveq6} becomes nonlinear. If we make the substitution $\omega\rightarrow\omega+\delta\omega_{L}$ in Eq.~\ref{eq:caveq6}, we can separate the envelope equation into linear and nonlinear terms.

\begin{equation}
\begin{array}{l}
\frac{\dot{\mathbf{V}}_{\mathrm{cav}}}{\omega_{0}}+\left[\frac{\omega_{0}^{2}+\omega^{2}}{4Q_{L}\omega^{2}}+j\frac{\omega_{0}^{2}-\omega^{2}}{2\omega\omega_{0}}\right]\mathbf{V}_{\mathrm{cav}}- \\ \\
\left[\frac{\omega_{0}^{2}}{2Q_{L}\omega^{2}}+j\frac{\omega_{0}^{2}+\omega^{2}}{2\omega\omega_{0}}\right]\frac{\delta\omega_{L}}{\omega}\mathbf{V}_{\mathrm{cav}} = \frac{j\dot{\mathbf{F}}+\omega\mathbf{F}}{\omega{Q_{e}}}
\end{array}
\label{eq:loreq2}.
\end{equation}

In Eq.~\ref{eq:loreq2}, $\frac{\delta\omega_{L}}{\omega}\ll1$ for most scenarios, therefore the nonlinear terms are much smaller than the linear terms.

\subsubsection{Pressure detuning}
Low temperature SRF structures are cooled in liquid Helium (LHe). During a quench, the cavity surface becomes normal conducting and power is dissipated through Ohmic losses and converted into heat. The heat in turn boils the surrounding Helium, resulting in an increased pressure exerted on the cavity.

A quench typically starts at a single point and quickly spreads over the surface of the cavity. As the normal conducting region grows, the heat produced from Ohmic losses increases rapidly and causes the LHe to violently flash-boil on the outer surface of the cavity. The volumetric expansion ratio for liquid to gaseous Helium is 1:757; thus when the LHe boils, it results in a sudden pressure increase.

If the LHe temperature is above 2.17$^{\circ}$K, the liquid is a normal fluid and has a viscosity. The rapidly expanding volume of gas pushes again the viscous liquid and causes an increased pressure on the cavity wall. If the LHe temperature is below 2.17$^{\circ}$K, the liquid is superfluid and has no viscosity. The boiling Helium experiences no resistance from the LHe and the pressure increase is caused by the increase in vacuum pressure from the gaseous Helium. During a quench, pressures are expected to increase up to $10^{2}-10^{3}$~mbar as the LHe boils in a normal fluid state because of the reaction force of the rapidly expanding gas on the viscous fluid. As superfluid LHe boils, the pressure increase is substantially reduced as the gaseous Helium expands into the vacuum, thus the pressure increase is expected to be $\sim1-100$~mbar. Measurements of frequency shift vs pressure range from 0.1-55~Hz/mbar~\cite{Neumann:2010zz,fermilab}. Therefore we expect the frequency shift due to pressure detuning during a quench to be of the order of kHz for LHe above 2.17$^{\circ}$K and of the order of $10^{0}-10^{2}$~Hz for superfluid Helium.

A physically realistic model of pressure detuning is difficult as there are many unknowns, such as the location of the quench point. These unknowns affect the transition time of the quench, the pressure distribution around the cavity and therefore the actual frequency shift due to pressure. We have assumed that the frequency shift due to pressure detuning is of the form

\begin{equation}
\delta{f}_{P}=\frac{\Delta{f_{p}}}{2}\left(1+\tanh{\left(\frac{t-t_{quench}}{\tau_{p}}\right)}\right)
	\label{eq:preseq1}
\end{equation}

\noindent where $\Delta{f_{p}}$ is the maximum frequency shift, $t_{quench}$ is the start time of the quench and $\tau_{p}$ is the transition time. This model is not realistic because the pressure detuning begins before the quench occurs, violating causality, but this is negligible if $\tau_{p}$ is small. In addition, given that $\tau_{p}$ is small, the pressure detuning occurs rapidly and the exact form of pressure as a function of time is not important.

If $\tau_{p}$ is comparable or greater than the LLRF latency, the LLRF system can react and compensate for the detuning, which will result in a short oscillatory response until the pressure detuning stabilises.

\subsubsection{Microphonics and mechanical oscillations}

Microphonics is the term used to describe the periodic frequency shift due to vibrations caused by nearby equipment, such as the cryogenic system and cooling pumps, as well as the impulse from the two previously described mechanical detuning mechanisms. These vibrations are broadband and allow the SRF structure to oscillate at its mechanical resonant frequency~\cite{delayen2006ponderomotive,kyrre2016}. Assuming that the SRF structure only has one mechanical resonant frequency, we can model the microphonic detuning as

\begin{equation}
\delta{f}_{m}=\Delta{f_{m}}\sin{\left(\omega_{m}t+\phi\right)}
	\label{eq:miceq1}
\end{equation}

\noindent where $\Delta{f_{m}}$ is the amplitude of the microphonic frequency shift, $\omega_{m}$ the mechanical resonant angular frequency of the SRF structure and $\phi$ and arbitrary phase shift. We can extend Eq.~\ref{eq:miceq1} to include additional terms of different resonant frequencies if desired.

Microphonics, pressure and Lorentz force detuning are all mechanical detuning mechanisms. Therefore we can expect all three mechanisms to excite oscillations. We can define the total frequency shift, $\delta{f}_{tot}$, due to all detuning mechanisms as

\begin{equation}
\delta{f}_{tot}=\delta{f}_{R}+\delta{f}_{mech}
	\label{eq:miceq2}
\end{equation}

\noindent where $\delta{f}_{mech}$ is the frequency shift due to the mechanical detuning mechanisms. $\delta{f}_{mech}$ is the solution for a driven, damped harmonic oscillator where $\delta{f}_{L}$, $\delta{f}_{p}$ and $\delta{f}_{m}$ are the driving terms. A simple model can be described as~\cite{kyrre2016}

\begin{equation}
\ddot{\delta{f}}_{mech}+2\eta\omega_{m}\dot{\delta{f}}_{mech}+\omega_{m}^{2}\delta{f}_{mech} = \ddot{\delta{f}}_{dr}
	\label{eq:miceq3}
\end{equation}

\noindent where $\eta$ is the mechanical damping coefficient and $\delta{f}_{dr}$ is the driving term given as 

\begin{equation}
\delta{f}_{dr} = \left(\delta{f}_{L}+\delta{f}_{p}+\delta{f}_{m}\right)e^{-j\omega_{m}\left(t-t_{quench}\right)}
	\label{eq:miceq4}
\end{equation}

It is assumed that the damping term is small and can be neglected when simulated over small timescales. The mechanical resonant frequency of the RF structure is of the order of $10^{0}$-$10^{4}$~Hz, however the pressure detuning occurs on the scale of MHz. We shall assume that $\delta{f}_{mech}$ from the pressure detuning only will have a form

\begin{equation}
\delta{f}_{mech}=\frac{\Delta{f_{p}}}{2}\left(1+\tanh{\left(\frac{\delta{t}}{\tau_{p}}\right)}\right)\left(1+je^{-j\omega_{m}\delta{t}}\right)
	\label{eq:miceq5},
\end{equation}

\noindent where $\delta{t}=t-t_{quench}$. If we determine the second derivative and rearrange as well as taking the approximation $1/\tau_{p}\gg\omega_{m}$, we obtain

\begin{equation}
\begin{array}{l}
\ddot{\delta{f}}_{mech}+\left[\omega_{m}^{2}+\frac{2}{\tau^{2}_{p}}\sech^{2}{\left(\frac{\delta{t}}{\tau_{p}}\right)}\right]\delta{f}_{mech} \\ \\
 = \omega^{2}_{m}\delta{f}_{p}+\frac{\Delta{f}_{p}}{\tau^{2}_{p}}\sech^{2}{\left(\frac{\delta{t}}{\tau_{p}}\right)}\left(1+e^{-j\omega_{m}t}\right)
\end{array}
	\label{eq:miceq6},
\end{equation}

\noindent where we see that the resonant frequency of mechanical oscillation becomes $\omega_{m}^{2}+\frac{2}{\tau_{p}}\sech^{2}{\left(\frac{\delta{t}}{\tau_{p}}\right)}$. This implies that during the rapid change in Helium pressure, the mechanical frequency briefly increases up to $\frac{\sqrt{2}}{\tau_{p}}$, allowing the RF frequency to rapidly change.

In simulation, the frequency shift due to pressure detuning is determined using Eq.~\ref{eq:miceq5} rather than evaluating Eq.~\ref{eq:miceq6} because the result is the same, but requires fewer computational steps. Similarly, microphonics are modeled as in Eq.~\ref{eq:miceq1}. The frequency shift due to Lorentz detuning is determined by evaluating

\begin{equation}
\ddot{\delta{f}}+\omega_{m}^{2}\delta{f}=\omega^{2}_{m}\delta{f}_{L}
	\label{eq:miceq7},
\end{equation}

\noindent because the detuning depends on the cavity RF voltage and must therefore be evaluated in parallel with the RF voltage. Since Lorentz detuning occurs over much longer timescales than the RF period, an RK4 integrator is unnecessary and a forward Euler integrator is used instead. For each time step, the frequency shift due to Lorentz detuning is calculated as

\begin{equation}
\begin{array}{l}
\dot{\delta{f}}_{n} = \dot{\delta{f}}_{n-1}+\omega_{m}^{2}\left(\delta{f}_{L}-\delta{f}_{n-1}\right)\delta{t} \\ \\
\delta{f}_{n} = \delta{f}_{n-1}+\dot{\delta{f}}_{n}\delta{t}
\end{array}
	\label{eq:miceq8}.
\end{equation}

\subsection{Low-level RF system}

In the model of the RF cavity and the detuning mechanisms, described in the previous sub-sections, Lorentz detuning is the only source of nonlinearities in our modelled system. As explained in the previous sub-section, the nonlinear terms from Lorentz detuning are several orders of magnitude smaller than the linear terms and can therefore be neglected. As such, there is no need to simulate a sophisticated LLRF system, capable of nonlinear corrections. Hence, the LLRF system is assumed to be a proportional-integral (PI) controller~\cite{Baudrenghien,Baudrenghien2,dexter}.

The cavity RF voltage is measured at regular intervals, given by the digital sampling rate; for SPS and HL-LHC simulations this is taken to be 40~MHz. The cavity RF in-phase ($I$) and quadrature ($Q$) voltages are measured relative to a reference clock signal. This is effectively the real and imaginary components of the cavity RF relative to the nominal RF frequency, and therefore provides all the relevant information about the cavity RF voltage. The PI controller compares the measured $I$ and $Q$ voltages to the set values stored and calculates a correction.

The correction to the cavity RF from the PI controller consists of two terms; a proportional and an integral term. The proportional term calculates a correction based on the previous measurement of the $I$ and $Q$ voltage, and corrects for fast changes in the cavity voltage. The integral term calculates a correction based on the integral (or sum) of the $I$ and $Q$ voltage over all previous times and corrects for slower changes and analogue drifts. The PI controller response is given as

\begin{equation}
\delta\mathbf{F}=c_{p}\left(\mathbf{V}_{\mathrm{cav}}-\mathbf{V}_{0}\right)+c_{i}\sum_{i}{\left(\mathbf{V}_{\mathrm{cav}}-\mathbf{V}_{0}\right)}
	\label{eq:llrf1},
\end{equation}

\noindent where $c_{p}$ and $c_{i}$ are the gains for the proportional and integral controllers respectively. The LLRF has a latency, which is typically much longer than the digital sampling rate; for the following simulations it is taken to be 1~{\textmu}s.

\subsection{Beam loading}

As a charged particle bunch travels through an RF structure, the particles interact with the RF fields. In the case of an accelerating cavity, if the bunch is accelerated, energy has been taken from the RF cavity. The stored energy in a cavity is related to the voltage as

\begin{equation}
U_{\text{stored}}=\frac{V_{\mathrm{cav}}^{2}}{\omega\left(\frac{R}{Q}\right)}
	\label{eq:bl1}.
\end{equation}

Thus, a change in stored energy will result in a change in voltage. For a deflecting or crabbing cavity, the energy gain of the particle bunch depends on the transverse position of the bunch as it passes through the cavity. The Panofsky-Wenzel theorem~\cite{pw} relates the transverse voltage, $V_{\bot}$, to the transverse variation of the longitudinal electric field

\begin{equation}
\mathbf{V}_{\bot}=-\frac{jc}{\omega}\int{\nabla_{\bot}\mathbf{E}_{z}\cdot{dz}}=-\frac{jc}{\omega}\nabla_{\bot}\mathbf{V}_{z}
	\label{eq:bl2}.
\end{equation}

In these simulations, it is assumed that the cavity operates in a dipole mode~\cite{PhysRevSTAB.17.104001,Brett201479,Calaga:2012zza,garcia} and that $\mathbf{V}_{z}\left(x\right)$ varies linearly with $x$ and $\mathbf{V}_{z}\left(0\right)=0$; where $x$ is taken to be the direction of the deflecting RF field. Therefore the transverse voltage is

\begin{equation}
\mathbf{V}_{\bot}=-\frac{jc}{\omega}\frac{\mathbf{V}_{z}\left(x\right)}{x}
	\label{eq:bl3}.
\end{equation}

Eq.~\ref{eq:bl1} is valid for both transverse and longitudinal voltages, provided the appropriate definition of $R/Q$ is used. The transverse $R/Q$ can be defined by substituting Eq.~\ref{eq:bl3} into Eq.~\ref{eq:bl1}

\begin{equation}
\left(\frac{R}{Q}\right)_{\bot}=\frac{V_{z}^{2}\left(x\right)}{\omega\left(\frac{x\omega}{c}\right)^{2}U_{\text{stored}}}
	\label{eq:bl4}.
\end{equation}

By conservation of energy, the change in stored energy, $\delta{U_{\text{stored}}}=qV_{beam}$, and the energy gain from the bunch is $-qV_{beam}$ where $q$ is the bunch charge and $V_{beam}$ is the voltage seen by the beam. The negative sign for the energy gain of the bunch is because a positive voltage accelerates a negatively charged particle. Assuming the change in stored energy is small, we can determine the change in cavity voltage from Eq.~\ref{eq:bl1}

\begin{equation}
\delta{U_{\text{stored}}}=qV_{beam}=\frac{2V_{\bot}}{\omega\left(\frac{R}{Q}\right)_{\bot}}\delta{V_{\bot}}
	\label{eq:bl5}.
\end{equation}

If we assume that the bunch passes through the cavity with a phase error, $\phi$, then $V_{beam}=V_{z}\left(x\right)e^{j\phi}$, where $x$ is the transverse position of the bunch centroid. Therefore we can determine the instantaneous change in cavity voltage by sustituting Eq.~\ref{eq:bl3} into Eq.~\ref{eq:bl5} as

\begin{equation}
\delta{\mathbf{V}_{\bot}}=\frac{q\omega}{2}e^{j\phi}\left(\frac{\omega{x}}{c}\right)\left(\frac{R}{Q}\right)_{\bot}
	\label{eq:bl6}.
\end{equation}

Note that the change in voltage depends on the transverse and longitudinal position of the bunch. Therefore the cavity voltage depends on the beam dynamics of the bunch, conversely, the beam dynamics depends on the state of the cavity. Hence there is a complicated interaction between the cavity and the beam. However, from initial studies described later in this article, we conclude that the LLRF system is easily able to compensate for beam loading, which allows us to neglect the effect the beam has on the cavity fields; significantly simplifying the simulations.

\section{Crab cavity quench measurements}
To verify the model of the quench dynamics and its impact on the crab cavity RF system, measurements of the RF amplitude and phase in order to determine the frequency shift during quench were taken at SM18~\cite{sm18}, CERN, on the Double Quarter-Wave (DQW) Proof-of-Principle (PoP) crab cavity. The measurements were performed in a vertical test configuration where the cavity is immersed in a liquid Helium bath, and the external Q is chosen to be slightly lower than the intrinsic Q-factor to minimise power requirements. In order to mitigate microphonics the measurement was run as a self-excited loop (SEL)~\cite{SelfExcitedLoops} where the drive frequency is locked to the cavity frequency. This configuration is not an exact replica of the setup in HL-LHC as the drive frequency will be fixed by a master oscillator for HL-LHC, and the loaded Q factor will be orders of magnitude less, but from these measurements it is possible to reconstruct the change in cavity frequency with time in order to apply this to the LLRF system model. In addition the Lorentz force detuning and pressure stability of the bare PoP cavity, with a stiffening frame, are significantly higher than in the dressed DQW cavity due to the LHe vessel, hence the frequency shift contributions will need to be separated and scaled accordingly.

The cavity is configured with a fundamental power coupler (FPC), with an external Q slightly less than the cavity ohmic Q prior to the quench, and a pick-up probe set to have an external Q factor at least an order of magnitude above the ohmic Q prior to the quench. The pick-up probe allows a transmitted power to be obtained to give a direct measurement of the stored energy in the cavity, given the external Q is measured during the calibration of the experiment, and also as a feedback signal to the SEL. Power is delivered to the cavity via the FPC, and we measure the input power and any power reflected back towards the source.

\subsection{Stored energy and transverse voltage}
To measure the cavity amplitude and phase we down-mix the input and output waveforms with a local oscillator operating at a frequency of around 400 MHz, set to be close to the cavity frequency prior to the quench. From the experimental measurements, we obtain the amplitudes of forward ($P_{F}$), reflected ($P_{R}$) and transmitted ($P_{T}$) power, as well as the respective RF phases relative to local oscillator signal ($\phi_{F}$, $\phi_{R}$ and $\phi_{T}$). In addition, the fundamental power coupler is measured to have a Q-factor, $Q_{in}=1.2\times10^{9}$, and the pickup probe to measure the transmitted power has a Q-factor, $Q_{T}=1.5\times10^{11}$. During a quench, the cavity RF system is not in equilibrium, so from the conservation of energy, where the drive is locked to the cavity frequency, we obtain

\begin{equation}
P_{F} - P_{R} - P_{T} - P_{C} - \frac{dU}{dt} = 0
\label{eq:meas1}
\end{equation}

\noindent where $P_{C}$ is the power dissipated in the cavity and $U$ is the stored energy in the cavity. When the RF system and the cavity are in equilibrium, $dU/dt=0$. From the definition of Q-factor, we can relate the stored energy, $U$, to $Q_{T}$ and $P_{T}$ as

\begin{equation}
U=\frac{2Q_{T}P_{T}}{\omega}
\label{eq:meas2}.
\end{equation}

\begin{figure}
   \includegraphics*[width=82.5mm]{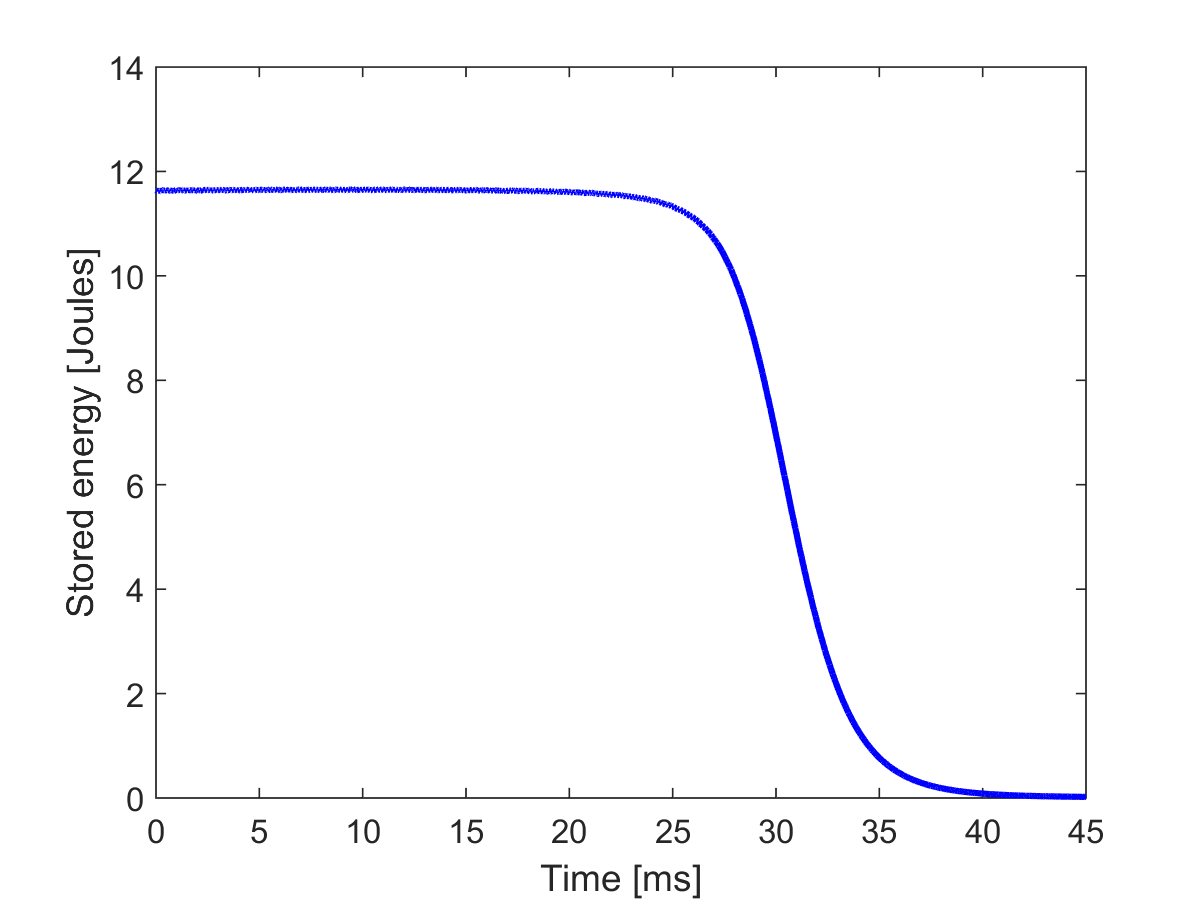}
   \caption{Stored energy vs time in the DQW PoP crab cavity during a quench measurement at SM18, CERN.}
   \label{fig:storedE}
\end{figure}

Figure~\ref{fig:storedE} shows the stored energy in the cavity, using Eq.~\ref{eq:meas2} and the measured data.

\begin{figure}
   \includegraphics*[width=82.5mm]{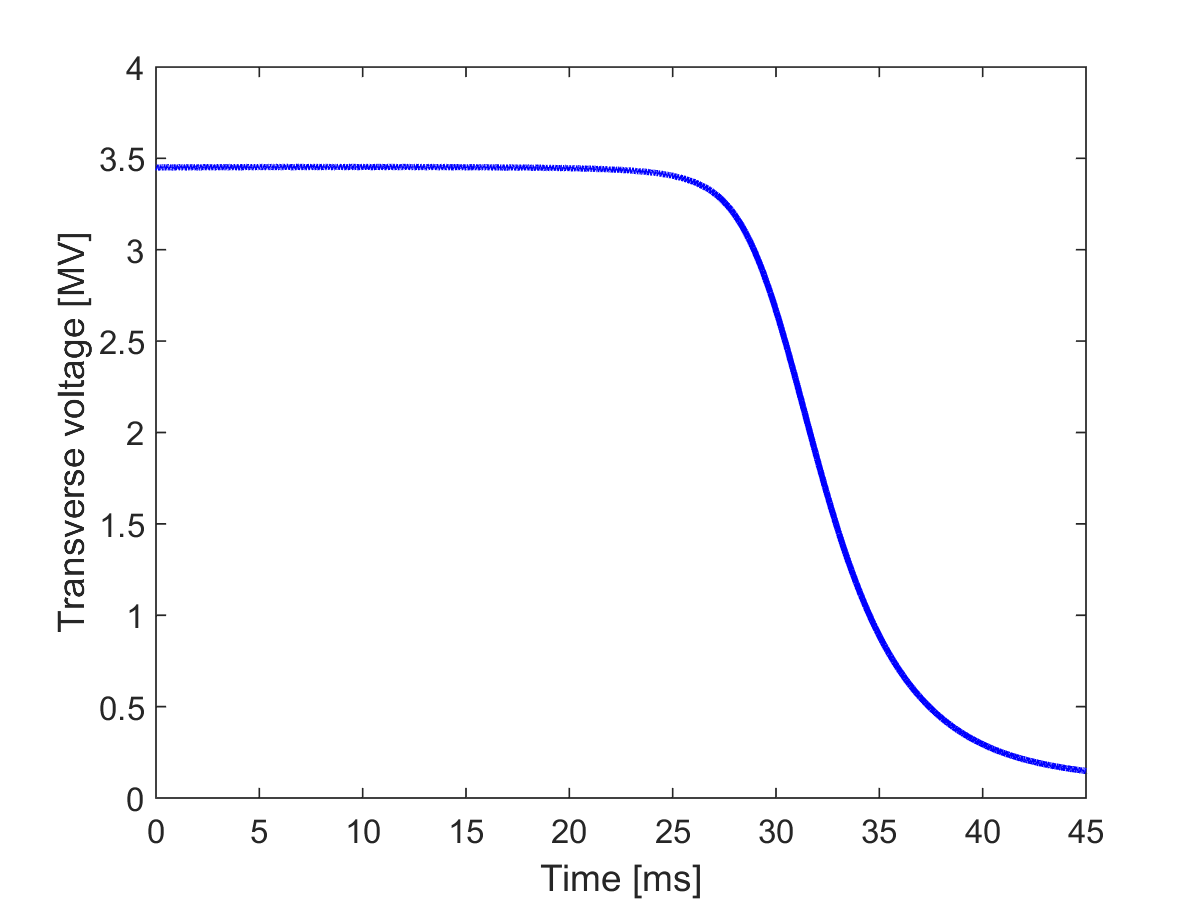}
   \caption{Transverse deflecting voltage, $V_{T}$ vs time in the DQW PoP crab cavity during a quench measurement at SM18, CERN.}
   \label{fig:VT_meas}
\end{figure}

\noindent The stored energy in the crab cavity can be related to the transverse deflecting voltage as

\begin{equation}
V_{T}^{2}=kU
\label{eq:meas3},
\end{equation}

\noindent where $k$ has been experimentally measured previously~\cite{PhysRevSTAB.18.041004} to be 1.0234~MV$^{2}$/J for the DQW PoP crab cavity. The measured transverse deflecting voltage is shown in Figure~\ref{fig:VT_meas}.

\subsection{Intrinsic Q-factor}
The intrinsic Q-factor, $Q_{0}$, can be related to the stored energy in the cavity and the power dissipated in the cavity as $Q_{0}=\frac{\omega{U}}{P_{C}}$. From Eqs.~\ref{eq:meas1} and \ref{eq:meas2}, we can define $Q_{0}$ as

\begin{equation}
Q_{0} = \frac{Q_{T}P_{T}}{P_{F} - P_{R} - P_{T} - \frac{dU}{dt}}
\label{eq:meas4},
\end{equation}

\begin{figure}
   \includegraphics*[width=82.5mm]{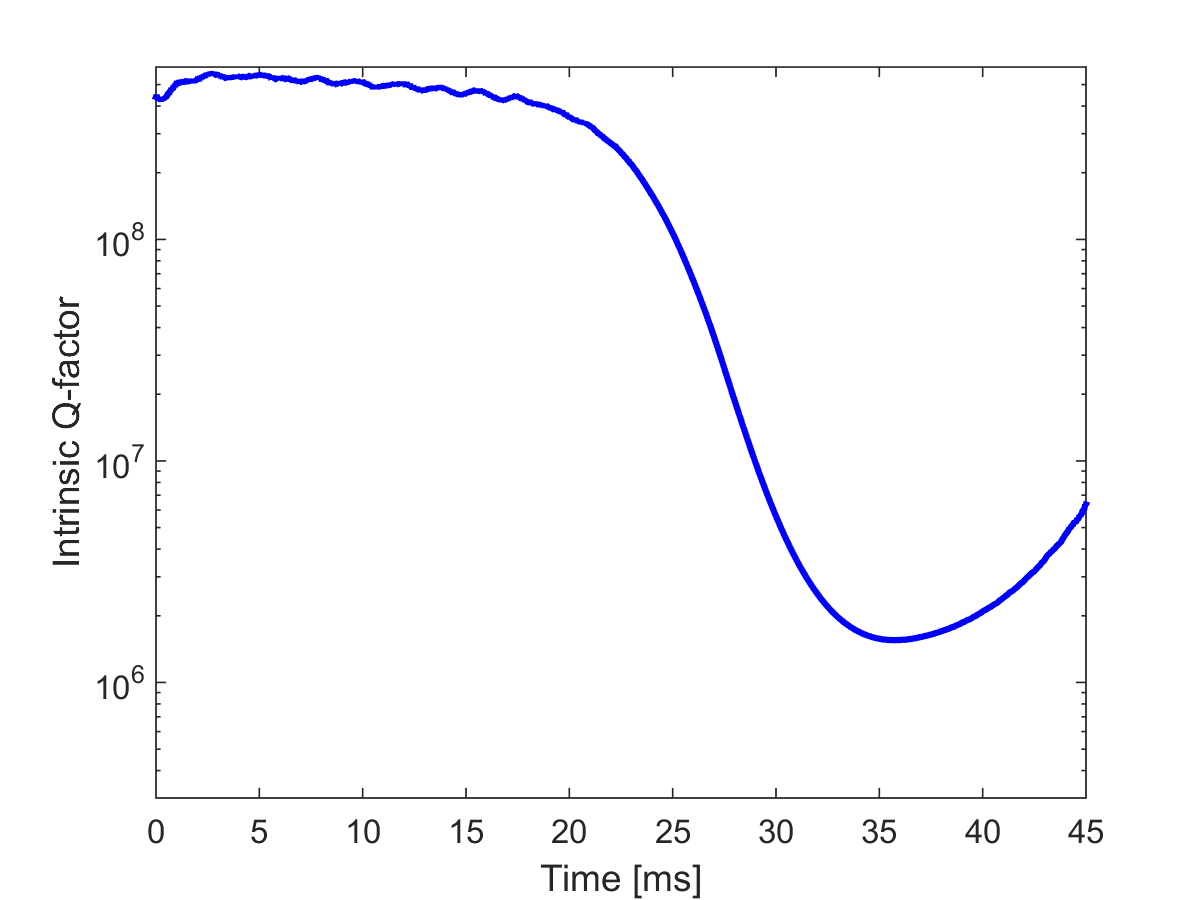}
   \caption{Intrinsic Q-factor, $Q_{0}$, vs time in the DQW PoP crab cavity during a quench measurement at SM18, CERN.}
   \label{fig:Q0_meas}
\end{figure}

\noindent $Q_{0}$ vs time is shown in Figure~\ref{fig:Q0_meas}. The vertical axis is logarithmic because the Q-factor drops by more than two orders of magnitude over the duration of the quench. The Q-factor begins to increase again from about 35~ms because the measured data shows a `soft' quench where the cavity recovers before the cavity becomes fully normal conducting. By comparison, in the simulations we present later in this paper, we only model `hard' quenches.

For the model of quench dynamics, we assume that the intrinsic Q-factor of the cavity decays exponentially during a quench as

\begin{equation}
Q_{0}\left(t\right) = Q_{0,NC} + \left(Q_{0,SC} - Q_{0,NC}\right)e^{-\frac{t}{\tau_{q}}}
\label{eq:meas5},
\end{equation}

\noindent where $Q_{0,SC}$ and $Q_{0,NC}$ are the superconducting and normal-conducting $Q_{0}$ respectively and $\tau_{q}$ is the quench transition time. If we assume $Q_{0,NC}\ll{Q_{0,SC}}$, then we can obtain

\begin{equation}
\ln\left(Q_{0}\left(t\right)\right) \approx \ln\left(Q_{0,SC} - Q_{0,NC}\right) - \frac{t}{\tau_{q}}
\label{eq:meas6}.
\end{equation}

\noindent By fitting a straight line to $\ln\left(Q_{0}\right)$ vs time during the quench transition (25~ms$~\leq{t}\leq30$~ms in Figure~\ref{fig:Q0_meas}), we estimate the quench transition time to be $\tau_{q}=1.63$~ms.

\subsection{Frequency shift}
\begin{figure}
   \includegraphics*[width=82.5mm]{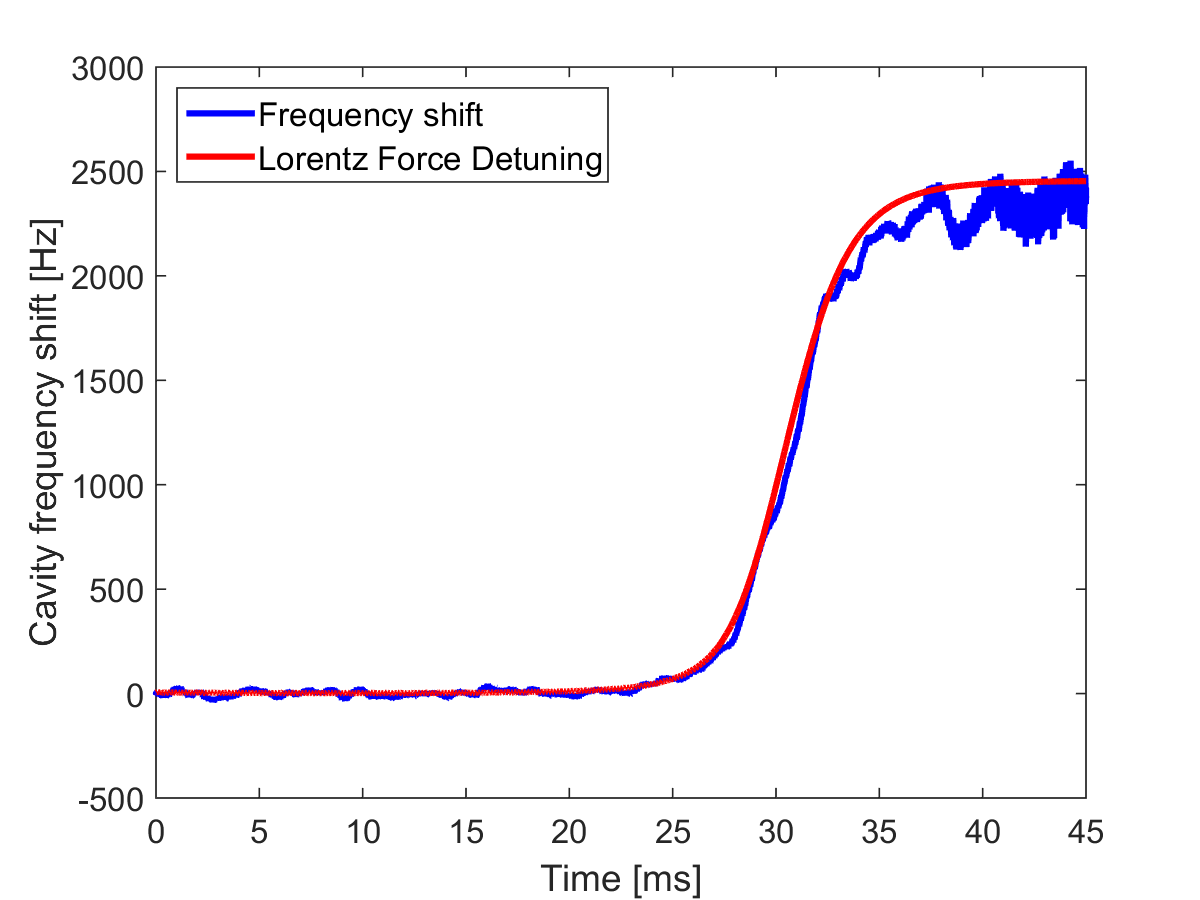}
   \caption{Frequency shift, $\delta{f_{\mathrm{cav}}}$ (blue), and calculated Lorentz Force detuning (red) vs time in the DQW PoP crab cavity during a quench measurement at SM18, CERN.}
   \label{fig:freq_shift}
\end{figure}

From the power amplitudes, we have been able to determine $U$, $V_{T}$ and $Q_{0}$ vs time. From the phase measurement, we are able to determine the frequency shift of the cavity during a quench. The frequency shift can be calculated from any of the phase data, but we have chosen to use $\phi_{T}$. Given that $\phi_{T}$ is measured in degrees, the frequency shift is given as

\begin{equation}
f_{cav,0} + \delta{f_{\mathrm{cav}}} = \frac{1}{360}\frac{d\phi_{T}}{dt} + f_{LO}
\label{eq:meas7},
\end{equation}

\noindent where $f_{cav,0}$ is the cavity resonant frequency before the quench, $\delta{f_{\mathrm{cav}}}$ is the change in cavity frequency over time and $f_{LO}$ is the local oscillator frequency. Figure~\ref{fig:freq_shift} shows the frequency shift, $\delta{f_{\mathrm{cav}}}$ (blue), vs time. As can be seen the frequency shift measured is very close to the calculated detuning from the stored energy and the simulated detuning factor. This provides confidence that the RF system is indeed frequency locked to the cavity frequency. In the measurement data before the quench, the cavity resonant frequency is measured to be 3.005~kHz higher than the local oscillator frequency; this has been subtracted from the data to only show the change in frequency during the quench. From Eq.~\ref{eq:loreq1}, we can calculate the frequency shift due to Lorentz Force detuning (red), given that the Lorentz detuning factor is measured to be $K_{L}=-206$~Hz/MV$^{2}$~\cite{PhysRevSTAB.18.041004} for the DQW PoP crab cavity.

\begin{figure}
   \includegraphics*[width=82.5mm]{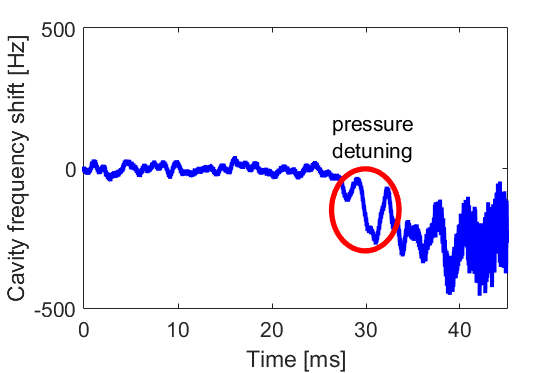}
   \caption{Frequency shift with Lorentz Force detuning subtracted vs time in the DQW PoP crab cavity during a quench measurement at SM18, CERN.}
   \label{fig:freq_shift_no_lorentz}
\end{figure}

By plotting the frequency shift with the predicted Lorentz Force detuning subtracted (Figure~\ref{fig:freq_shift_no_lorentz}), we are able to observe the effect of pressure detuning as well as mechanical vibrations in the cavity at several different frequencies, although we cannot be certain that the oscillations later in the pulse when the transmitted power is low are not due to self-excited loop (SEL) instabilities. The effect of the pressure detuning is shown by the red circle. The slow drop in frequency after 20~ms is due to a low frequency mechanical vibration (10-20~Hz) and a higher frequency mechanical oscillation ($\sim200$~Hz) is also visible. The DQW PoP crab cavity's resonant frequency has a measured pressure sensitivity of 33.6~Hz/mbar~\cite{PhysRevSTAB.18.041004}. The observed pressure detuning during a quench of the DQW PoP cavity is 100-200~Hz.

\subsection{Extrapolation to HL-LHC}
As previously stated, the DQW PoP crab cavity has a measured Lorentz Force detuning factor of -206~Hz/MV$^{2}$ and a pressure sensitivity of 33.6~Hz/mbar. By comparison, the dressed crab cavity that will be installed in SPS has a measured Lorentz Force detuning factor of -40~Hz/MV$^{2}$ and a pressure detuning sensitivity factor of -0.103~Hz/mbar~\cite{DQWmeas}. These values are also expected to be the same for HL-LHC.

Based on these measured values and the measured frequency shift for the PoP cavity, we estimate a frequency shift for SPS/HL-LHC due to Lorentz Force detuning of $\sim460$~Hz and a pressure detuning of $\sim0.3-0.6$~Hz. For the PoP crab cavity, it was observed that Lorentz Force detuning and its associated microphonics were the dominant mechanism of frequency detuning; for SPS/HL-LHC, it is expected to be even more dominant. 

\subsection{Choice of quench parameters for simulations}
In reality, the dynamics governing the behaviour of a quench is a very complex, multi-physics problem. As such, the quench parameters can vary drastically from one quench to another in the same structure. For example, it is known that the quench transition time can vary by several orders of magnitude, depending on the initial location of the quench, due to the local physical and thermal properties of the cavity as well as the cause of the quench. For HL-LHC, quenches which occur on the order of the bunch revolution period are considered to have the highest potential to result in significant beam loss because they will cause every bunch to be scattered onto different orbits; resulting in large betatron oscillations. For very fast quenches, only a few bunches are likely to be scattered, while all other bunches will remain on the reference orbit; thus only limited beam losses would be expected. Conversely, very slow quenches are also not a problem because feedback systems within the HL-LHC ring will correct the orbits of scattered bunches and any particle losses will occur over many turns, by which point the interlock system will have dumped the beam.

For the simulation studies described in the following sections, we choose the worst-case scenario for quench parameters and therefore choose a quench transition time of $\tau_{q}=10$~{\textmu}s, which results in the transverse deflecting voltage taking approximately 1 revolution period of the HL-LHC to go from nominal value to the post-quench value ($\sim1$~kV). We will use a Lorentz detuning factor which is 5 times larger than the expected value to allow a margin of safety in the simulations. Similarly we will also use a pressure detuning factor which is a factor of 5 larger as the detuning factor assumes a uniform pressure on the cavity surface, the effect of a non-uniform pressure is less predictable. For simulations at 4K, we use a significantly higher pressure detuning factor to provide a frequency shift of 4000~Hz, which is consistent with the observed phase shift during a quench of the KEKB crab cavity~\cite{KEKBcrab}.

\section{Beam dynamics studies}

Particle tracking simulations were undertaken in two different ways. First, a method of transporting particles using sector maps produced by MAD-X~\cite{madx} and a second method where the Matlab cavity simulation results are read into the tracking code SixTrack~\cite{SixTrackmanual}, in order to model the interaction between the cavity and the bunch train.

In these simulations, we assume that only one crab cavity has failed. In the cases where there are multiple crab cavities (such as for HL-LHC) we assume all the other cavities are ideal and do not have phase or amplitude jitter.

\subsection{Tracking with sector maps}
A map refers to a transformation which describes how the particle phase space distribution changes from one location in a beam line to a location further downstream. In circular machines, a one-turn map refers to the transformation of the phase space distribution when the bunch travels an entire revolution of the ring and a sector map refers to the transformation of the phase space over some part of the ring.

A first order map, often known as a transfer matrix, can be expressed in index notation as

\begin{equation}
x^{(1)}_{i} = \sum_{j}{R_{ij}x^{(0)}_{j}}
	\label{eq:track1},
\end{equation}

\noindent where the superscripts 0 and 1 refer to the initial and final coordinates respectively. For these studies, we consider the first and second order tensors, in which case, the transformation is written as

\begin{equation}
x^{(1)}_{i} = \sum_{j}{R_{ij}x^{(0)}_{j}}+\sum_{j}{\sum_{k}{T_{ijk}x^{(0)}_{j}x^{(0)}_{k}}}
	\label{eq:track2}.
\end{equation}

If we now use a second sector map to transform the beam from position `1' to position `2' and express in terms of $x^{(0)}$, we obtain

\begin{equation}
\begin{array}{l}
x^{(2)}_{i} = \sum_{j}{\sum_{k}{R^{(2)}_{ij}R^{(1)}_{jk}x^{(0)}_{k}}}+ \\ \\
\sum_{j}{\sum_{k}{T^{(2)}_{ijk}\left(\sum_{l}{R^{(1)}_{jl}x^{(0)}_{l}}\right)\left(\sum_{l}{R^{(1)}_{kl}x^{(0)}_{l}}\right)}}+ \\ \\
\sum_{j}{R^{(2)}_{ij}\left(\sum_{k}{\sum_{l}{T^{(1)}_{jkl}x^{(0)}_{k}x^{(0)}_{l}}}\right)}+\mathcal{O}(x^{4})
\end{array}
	\label{eq:track3}
\end{equation}

Note in Eq.~\ref{eq:track3} that tracking through multiple second (or higher) order maps introduces additional higher order terms. In the sector map tracking, we truncate to second order as these higher order terms are orders of magnitude smaller than the first and second order terms and storing these higher order terms will significantly increase the required computing time to track particles over many turns.

The particles are tracked from one crab cavity to the next (or to the interaction points IP1 and IP5) and each cavity is modeled as a thin kick by applying the angular deflection from the cavity to the beam at the longitudinal midpoint of the cavity. The tracking simulations using sector maps shown in this article track 3 particles, which initially lie on the reference orbit, but are longitudinally at the head, centre and tail of the bunch.

The use of sector maps is advantageous because it allows the particles to be tracked very quickly by avoiding the need to model every element in the beam line. However, in a ring this technique does not allow us to accurately predict beam losses as the particles are simply mapped from one location to another in a discrete manner. Sector map tracking was used to qualitatively study the cavity behaviour and the interaction between the beam and the cavity.

\subsection{Particle tracking in SixTrack}
Particle tracking in SixTrack allows for a more detailed analysis of the beam dynamics and losses, but requires significantly more simulation time than the sector map method.
It works by tracking the particles element-by-element through the lattice of the machines.
Two versions of this code are available: The ``standard'' version that only considers the deterministic particle dynamics~\cite{SixTrackmanual,SixTrack_status,SixTrack_IPAC17}, and the ``collimation'' version~\cite{SixColl} which also incorporates Monte-Carlo routines for scattering the particles in the collimator jaws.
This scattering is important for understanding the spatial distribution of the losses, especially in the superconducting magnets.

In general, the ``standard'' SixTrack version was used as no collimation setup was available for the SPS; here the losses were computed by comparing $\{x,y\}$ position of each particle to the aperture at the aperture bottlenecks.
For the HL-LHC, where collimation input files are available~\cite{HL_coll,CollSoft,PhysRevSTAB.17.081004}, the collimation version of SixTrack was also run.
The latter code also takes the material of the bottlenecks (i.e.\ the primary collimators) into account, allows for scattering or absorption of the particles using a Monte-Carlo method, and computes a loss map along the ring.
A comparison between the losses code that only takes the bottlenecks into account, and the SixTrack collimation version was performed.
As shown in Figure~\ref{FIG:SixTrack:LHC:LossesTime}, the results are very similar, indicating that only considering the aperture bottlenecks is a reasonable approximation.

In the SixTrack simulations, each bunch in the bunch train is treated separately, and the cavity voltage and phase as a function of time are loaded from pre-calculated files using the Dynamic Kicks functionality (DYNK)~\cite{DYNK,DYNK2}, which updates the cavity parameters at the beginning of every turn.
The pre-calculated files were prepared as described earlier in this paper.

All the simulations were run for a total of 120 turns, where the cavity parameters are kept constant at their ``ideal'' values for the first 20 (SPS) or 100 (HL-LHC) turns, before the mentioned files are loaded. Note that the plots only show what happened after the cavity parameters were ``unlocked'' and allowed to vary.

In the SixTrack studies, we neglect beam loading. For HL-LHC, the bunch charge is approximately 18.4~nC and the RF amplifier for the cavity can deliver a maximum of 80~kW, from Eq.~\ref{eq:bl5}, we can estimate that for beam loading to be comparable to the maximum RF power, the full bunch train would need a transverse offset of approximately 4 mm. In reality, the expected offset is estimated to be of the order of 100~\textmu{m}. Furthermore, the assumption that the bunch train has a constant transverse offset is unphysical. In reality, the transverse motion of the beam undergoes betatron oscillations, which will cause the beam loading effect to cancel itself over approximately 3 revolutions of the ring.

For both the HL-LHC and SPS cases, the particles in each bunch were assumed to be distributed according to a correlated bivariate Gaussian with tails in each plane of the transverse phase space; the tails, representing 5\% of the total bunch population, were taken to have the same distribution and were 1.8 times wider in $\{x,x'\}$ and $\{y,y'\}$ as described in~\cite{Bruce:CCfail,Andrea15}.
The initial particle distribution was matched at the injection point to the transverse phase space using the given Twiss parameters, dispersion, orbit, and emittance at the injection point; for the transverse tails the phase space was scaled by the given factor.

For the longitudinal phase space, an uncorrelated bivariate Gaussian distribution was used; here the bunch length and energy spread were set to fill the bucket.
The energy spread was adjusted by hand in order to minimize the oscillation in bunch length, something that occurs when the distribution is not perfectly matched.
Some amount of this oscillation remained, however it is not likely to have any considerable impact on the results as the oscillation period (approximately $\frac{1}{2}$ the period of the synchrotron oscillations) is much slower than the processes under study.
In order to avoid simulating particles outside of the RF bucket, when generating the longitudinal distribution we check that the initial energy and $z$-position of each generated particle was inside the bucket, and generated new longitudinal coordinates for any particles that did not pass this test until they did pass; effectively cropping the distribution at the bucket edge with an accept/reject Monte-Carlo generator.

In total, each bunch was represented by 60'000 particles, 40'000 for the core and 20'000 for the tail; the finer sampling of the tails were used in order to improve the beam loss estimates.
Each bunch was generated from a separate random seed, and the bunches were different between the different machines but identical between different simulations of the same machine; i.e.\ the initial distribution of bunch 1 for HL-LHC was always the same, and so on for the rest of the bunches and for the SPS simulations.
This allows comparison between the simulations, subtracting the statistical noise from the effect of the crab cavity.

\begin{table}
  \caption{Key parameters of the SixTrack simulations for the three machines.}
  \label{TAB:SixTrack:MACHINEPARAMETERS}
  \begin{tabular}{l c | c c c }
    \hline
                     &   & \textbf{SPS / 55} & \textbf{SPS / 120} & \textbf{HL-LHC} \\
    \hline
    Optics name      &             & Nom.\ Q26   & Nom.\ Q26   & v1.2   \\
    Num.\ bunches    &             & 288         & 288         & 2810   \\
    Energy/$p^+$     &[GeV]        & 55          & 120         & 7000   \\
    Rev.\ time       &[{\textmu}s] & 9.7         & 9.7         & 88.9   \\
    $\epsilon_n$ H/V &[{\textmu}m] & 3.1 /2.8    & 3.1/2.8     & 2.5/2.5 \\
    Tune $Q_x$/$Q_y$ &             & 26.13/26.18 & 26.13/26.18 & 62.31/60.32 \\
    Sync.\ tune      &             & 0.00805     & 0.00583     & 0.00212 \\
    \hline
    CC plane         &             & V           & V           & V  \\
    $\beta_y$ at CC  &[m]          & 64.7        & 64.7        & 3760 \\
    \hline
    $\Delta y' = \frac{qV}{E}$  &[{\textmu}m] & 55          & 25          & 0.43 \\
    $\Delta y' /  \sqrt{\epsilon_g/\beta_y}$ & $[\sigma]$ & 2.02 & 1.36        & 1.97  \\
    \hline
  \end{tabular}
\end{table}

\section{Results}

\subsection{Benchmarking with KEKB results}
\begin{figure}
   \includegraphics*[width=82.5mm]{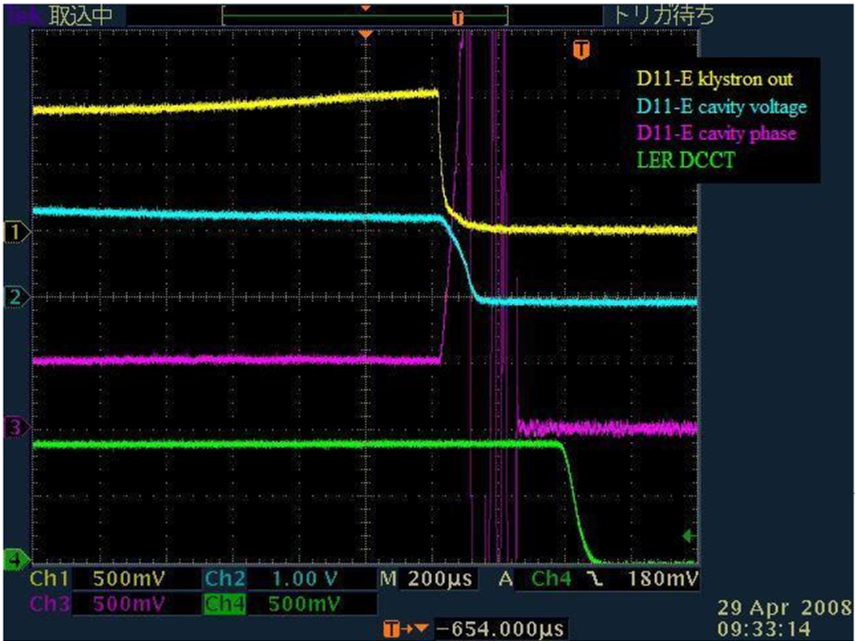}
   \caption{Oscilloscope screen shot showing klystron power (yellow), cavity voltage (cyan) and phase (magenta) and beam current (green) vs. time during a quench of a KEKB crab cavity~\cite{KEKBcrab}.}
   \label{fig:KEKBcav1}
\end{figure}

The motivation for this study was due to experimental measurements of the crab cavities installed at KEKB~\cite{KEKBcrab}. During a quench, KEKB observed the crab cavity phase shifting by up to 50$^{\circ}$ within 50~{\textmu}s as shown in Figure~\ref{fig:KEKBcav1}.

Figure~\ref{fig:KEKBcav1} shows that the klystron power begins to slowly increase approximately 800~{\textmu}s before the quench is detected and the RF is switched off. The beam is dumped approximately 400~{\textmu}s after the RF is switched off.

Due to the unpredictable nature of a quench, there are many variables, such as the location of the centre of the quench, which can greatly alter the transition time of the quench and the time at which pressure detuning becomes significant. Figure~\ref{fig:KEKBcav1} is an extreme example of a KEKB crab cavity quench, showing the maximum recorded rate of change of phase. It is likely that the pressure detuning rapidly increased at approximately the same time as the RF was switched off.

\begin{figure}
   \includegraphics*[width=82.5mm]{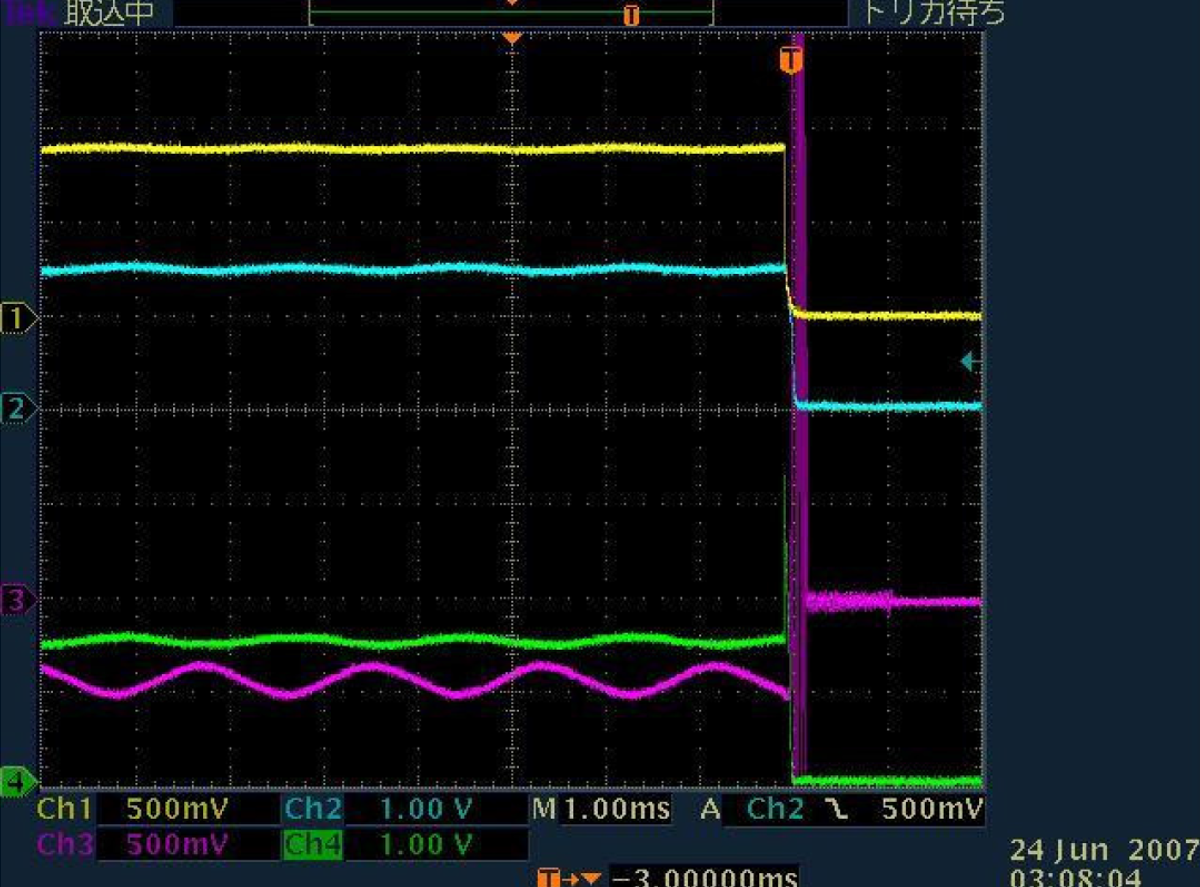}
   \caption{Oscilloscope screen shot showing microphonics and the resulting effect on tetrode power (yellow), cavity voltage (cyan) and phase (magenta) and beam current (green) vs. time~\cite{KEKBcrab}.}
   \label{fig:KEKBcav2}
\end{figure}

In addition to the observed behaviour due to a quench, microphonics were also observed at KEKB (Figure~\ref{fig:KEKBcav2}). Note that the microphonics affects the cavity phase more than the amplitude and that it predominantly consists of a single frequency component. This is generally true for SRF cavities because as shown in Eq.~\ref{eq:caveq6}, the RF phase is very sensitive to changes in frequency, but the amplitude is not.

\begin{figure}
  \centering
  \includegraphics*[width=85mm]{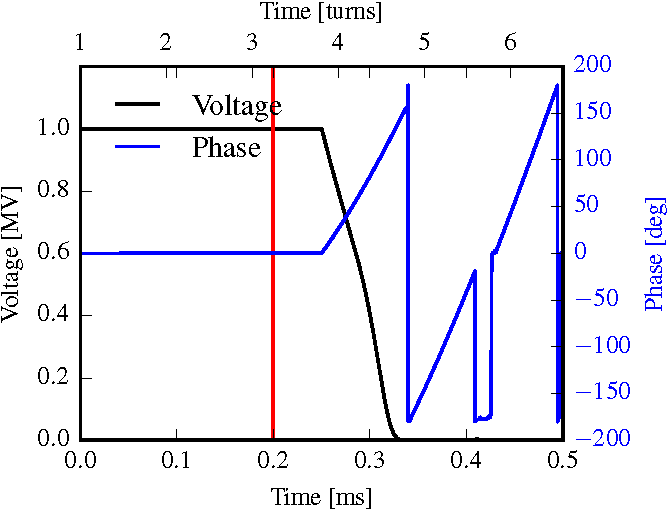}
  \caption{Plot of the simulated cavity voltage and phase vs.\ time during a quench for a KEKB-like cavity. The vertical red line represents the start time of the quench.}
  \label{fig:KEKBcav3}
\end{figure}

Figure~\ref{fig:KEKBcav3} shows the results of a simulation of a KEKB-like cavity during a quench. The simulation uses the HL-LHC beam optics, but the KEKB crab cavity parameters from~\cite{KEKBcrab}, summarised in Table~\ref{tab:ccparams}. Hence the beam loading will not accurately represent the true KEKB system; but these studies still allow a qualitative interpretation of some of the key features observed in Figure~\ref{fig:KEKBcav1}.

\begin{table}[hbt]
\centering
\caption{Comparison between the KEKB~\cite{KEKBcrab,kekb} and HL-LHC~\cite{Apollinari:2116337} Crab cavities}
\begin{tabular}{lcc}
\hline
\textbf{Cavity parameter} & \textbf{KEKB} & \textbf{HL-LHC} \\
\hline
Beam energy [GeV] & 8 & 7000 \\
Transverse voltage [MV] & 1 & 3 \\
Resonant frequency [MHz] & 509 & 400 \\
Transverse R/Q [$\Omega$] & 50 & 400 \\
Superconducting $Q_{0}$ & $10^{9}$ & $10^{9}$ \\
Normal conducting $Q_{0}$ & $10^{3}$ & $10^{3}$ \\
$Q_\mathrm{e}$ & $1\times10^{5}$ & $5\times10^{5}$ \\
Operating temperature [K] & 4 & 2 \\
Quench transition time [{\textmu}s] & 10 & 10 \\
\hline
\textbf{LLRF parameters} & & \\
\hline
Latency [{\textmu}s] & 1 & 1 \\
Digital refresh time [ns] & 25 & 25 \\
Proportional controller gain ($c_{p}$) & 6.06 & 30.3 \\
Integral controller gain ($c_{i}$) & 6.94$\times10^{-7}$ & 3.47$\times10^{-6}$ \\
Amplifier Q & 400 & 400 \\
Maximum tetrode power [kW] & 80 & 80 \\
Signal to noise ratio & 1000 & 1000 \\
\hline
\textbf{Detuning parameters} & & \\
\hline
Lorentz ($K_{L}$) [Hz/MV$^{2}$] & 200 & 200 \\
Pressure ($\Delta{f_{p}}$) at 4K [Hz] & 4000 & 4000 \\
Pressure ($\Delta{f_{p}}$) at 2K [Hz] & 100 & 100 \\
Microphonics ($\Delta{f_{m}}$) [Hz] & 2000 & 2000 \\
Mechanical frequency ($\omega_{m}$) [Hz] & 900 & 900 \\
\hline
\end{tabular}
\label{tab:ccparams}
\end{table}

In the simulation, the frequency shift due to pressure detuning, which depends on the stiffness and geometry of the cavity, was assumed to be $\Delta{f}_{p}=4$~kHz, but is consistent with measurements of other SRF structures, given that the boiling LHe causes a pressure increase on the cavity of 0.1-1~Bar~\cite{Neumann:2010zz, fermilab}. In Figure~\ref{fig:KEKBcav3}, the discrete jumps in phase between 0.4-0.5~ms are due to beam loading. As the cavity voltage is very low, the changing voltage due to the beam can shift the phase rapidly.

The results from the simulations show that when the RF is switched off, which is also assumed to be approximately when the pressure detuning becomes significant, the phase shifts by 80$^{\circ}$ in 50~{\textmu}s; similar to the 50$^{\circ}$ in 50~{\textmu}s measured at KEKB. In addition to the consistency of the rate of change of phase between simulation and measurement, the ramp down of the voltage amplitude is also consistent. The time for the voltage to ramp down is approximately 50~{\textmu}s for both cases and the `kink' on the curve is consistent; the kink is dependent on $Q_{L}$ and rate of change of $Q_{L}$ during a quench.

\begin{figure}
  \centering
  \includegraphics*[width=85mm]{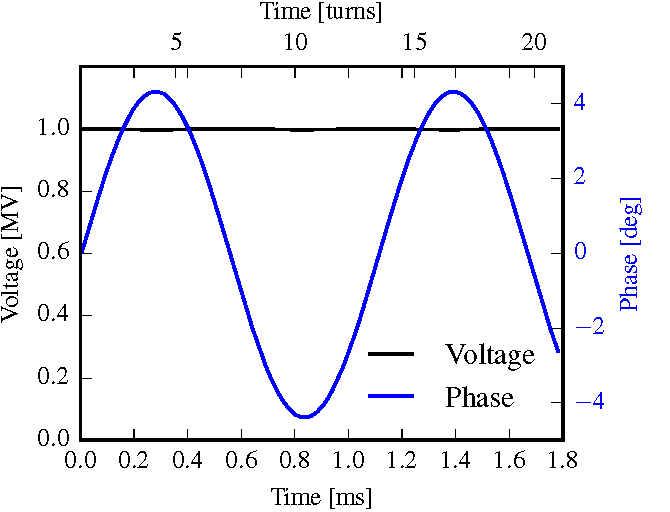}
  \caption{Plot of the simulated cavity voltage and phase vs.\ time under the influence of microphonics.}
  \label{fig:KEKBcav4}
\end{figure}

Figure~\ref{fig:KEKBcav4} shows a simulation of the microphonics in a KEKB-like cavity. The voltage amplitude varies by approximately 1.5\% and the time structure of the bunch train is observed due to beam loading. The frequency shift due to microphonics is assumed to be $\Delta{f}_{m}=2$~kHz and results in a phase oscillation of $\sim4.5^{\circ}$; the KEKB studies measured a 4$^{\circ}$ phase oscillation, showing good agreement.

\subsection{Parameter studies}
The benchmarking of the cavity quench model to measurements from the KEKB crab cavities has shown that the model can successfully reproduce the key features observed during a quench. However, this does not provide any insight into the factors which affect the cavity behaviour during a quench. In order to determine which physical phenomena affect the cavity, a study was undertaken that compares the simulation results when different effects were included or omitted in the model. For the simulations in this study, microphonics will be neglected in order to make the quench behaviour easier to see in plots.

Table~\ref{tab:ccparams} shows the cavity, LLRF and quench parameters used during the studies in this paper. For these studies, the quenching cavity is taken to be the crab cavity further upstream of interaction point, IP1 (ATLAS), for Beam~2. The bunch positions are taken at the centre of this cavity during tracking simulations.

\subsubsection{KEKB-like cavity}
\begin{figure}
   \includegraphics*[width=82.5mm]{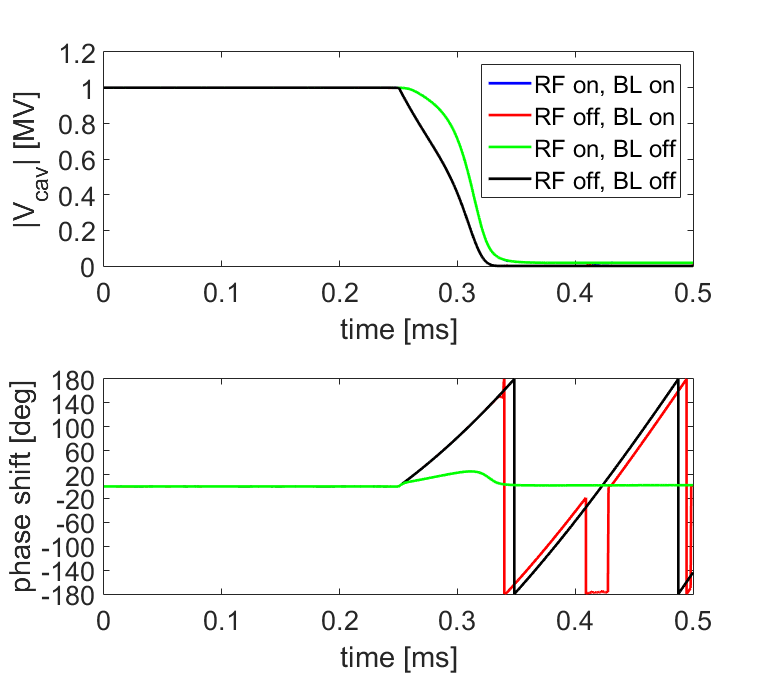}
   \caption{Plots of cavity voltage amplitude (top) and phase (bottom) for a KEKB-like cavity at 4K during a quench, comparing the effects of the RF system and beam loading.}
   \label{fig:paramcav1}
\end{figure}

Figure~\ref{fig:paramcav1} shows the amplitude and phase of the KEKB-like cavity when the LHe temperature is 4K. This is a comparison of the effects of including beam loading (BL) in the simulation (BL on/off) and keeping the RF on or off during a quench (RF on/off). Beam loading causes rapid fluctuations in the cavity voltage phase after the quench if the RF is switched off. Other than this, beam loading has an extremely small effect on the cavity voltage because the RF system is able to compensate.

\begin{figure}
   \includegraphics*[width=82.5mm]{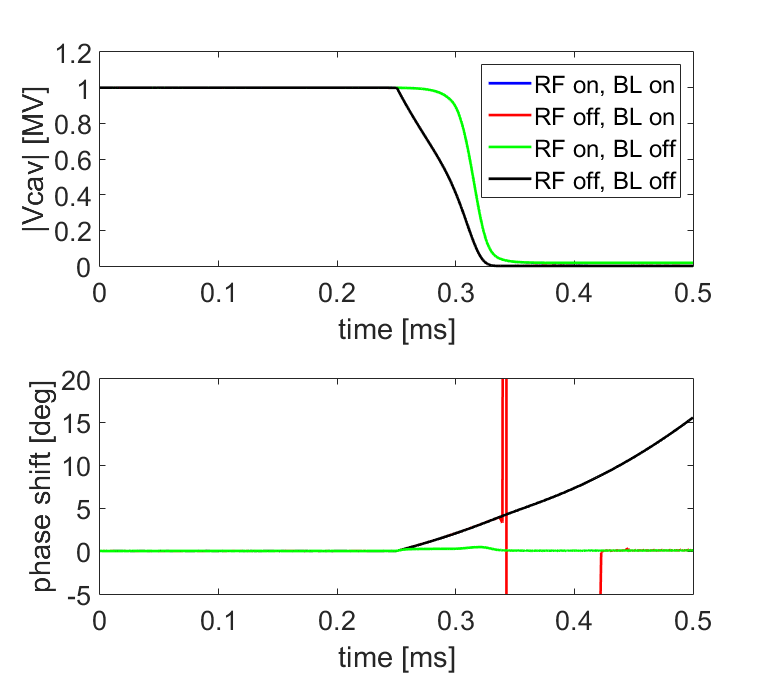}
   \caption{Plots of cavity voltage amplitude (top) and phase (bottom) for a KEKB-like cavity at 2K during a quench, comparing the effects of the RF system and beam loading.}
   \label{fig:paramcav2}
\end{figure}

In the phase plot in Figure~\ref{fig:paramcav1}, the case where beam loading is neglected and the RF turned off during the quench (black line) shows a discrete jump at $\sim0.9$~ms. This is because there is nothing to drive the cavity voltage and at this point the voltage is too small for Matlab and the phase becomes a default value depending on which quadrant of complex space the voltage occupies. It can be seen that the change in phase during a quench is reduced if the RF is kept on during a quench (green line). Figure~\ref{fig:paramcav2} shows the same comparison as Figure~\ref{fig:paramcav1}, but the LHe temperature is 2K and therefore superfluid.

\begin{figure}
   \includegraphics*[width=82.5mm]{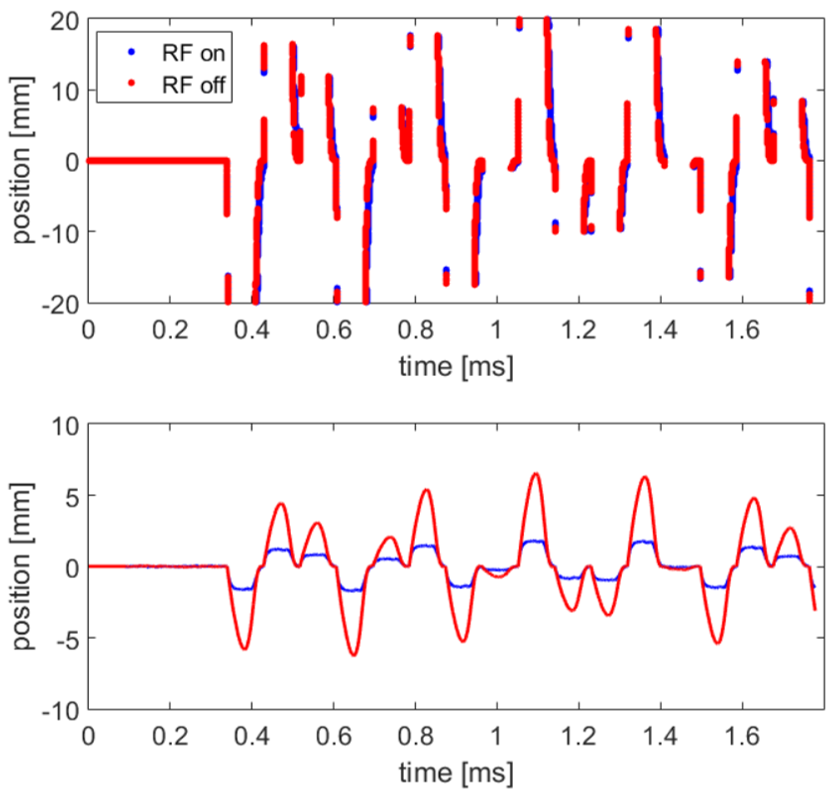}
   \caption{Transverse beam position during a quench in 4K LHe (top) and 2K LHe (bottom) for a KEKB-like cavity, with the RF kept on (blue) or switched off (red) when a quench is detected.}
   \label{fig:paramcav3}
\end{figure}

By comparing the phase plots in Figures~\ref{fig:paramcav1} and \ref{fig:paramcav2}, it is clear that the rapid phase shifts observed on the KEKB crab cavities is due to the LHe temperature and switching off the RF during the quench. Figure~\ref{fig:paramcav3} shows the transverse position of the centre of the bunches vs.\ time with the RF kept on (blue) or switched off (red) when the quench is detected. The top plot shows the comparison when the LHe temperature is 4K and the bottom plot when the LHe is at 2K. These plots use the sector map tracking method to provide a qualitative understanding.

In the top plot of Figure~\ref{fig:paramcav3}, particles are cut if the transverse position exceeds $\pm20$~mm; although this is not the actual aperture of the KEKB crab cavities, this is a reasonable aperture to assume for the optics used.

Table~\ref{tab:KEKtrans} shows the estimated number of particles which survive 10~turns after the quench for the KEKB case under different conditions. Only 3 particles are tracked per bunch for the sector map tracking, situated longitudinally at the head, middle and tail of the bunch. Thus the values in Table~\ref{tab:KEKtrans} are only meant to qualitatively understand the impact on the beam due to a crab cavity failure.

\begin{table}[bt]
\centering
\caption{Summary of fraction of particles surviving 10~turns after a quench of the KEKB-like cavity under different conditions.}
\begin{tabular}{lcc}
\hline
\textbf{Simulation} & \textbf{LHe at 2K} & \textbf{LHe at 4K} \\
\hline
RF on, BL on & 40.08\% & 8.42\% \\
RF off, BL on & 51.92\% & 9.76\% \\
RF on, BL off & 40.05\% & 8.43\% \\
RF off, BL off & 51.89\% & 9.77\% \\
\hline
\end{tabular}
\label{tab:KEKtrans}
\end{table}

As expected, beam loading causes a negligible effect on the fraction of the beam which survives 10~turns. However, counter-intuitively, keeping the RF on during a quench causes more particle losses than switching it off during a quench at both 2K and 4K. This can be explained by inspection of Figures~\ref{fig:paramcav1} and \ref{fig:paramcav2}. Although the RF reduces the phase shift during a quench, it also keeps the amplitude of the voltage higher for longer; hence providing more erroneous deflection to the bunch train for longer. Therefore for the KEKB case, it is preferable to protect the RF system and switch it off when a quench is detected.

\subsubsection{HL-LHC cavity}
\begin{figure}
\centering
   \includegraphics*[width=82.5mm]{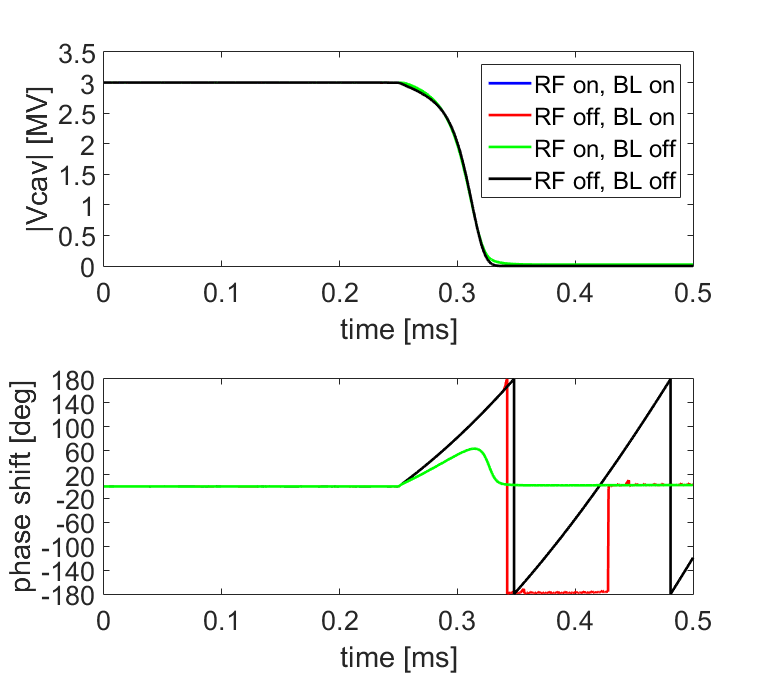}
   \caption{Plots of cavity voltage amplitude (top) and phase (bottom) for an HL-LHC cavity at 4K during a quench, comparing the effects of the RF system and beam loading.}
   \label{fig:paramcav4}
\end{figure}

\begin{figure}
\centering
   \includegraphics*[width=82.5mm]{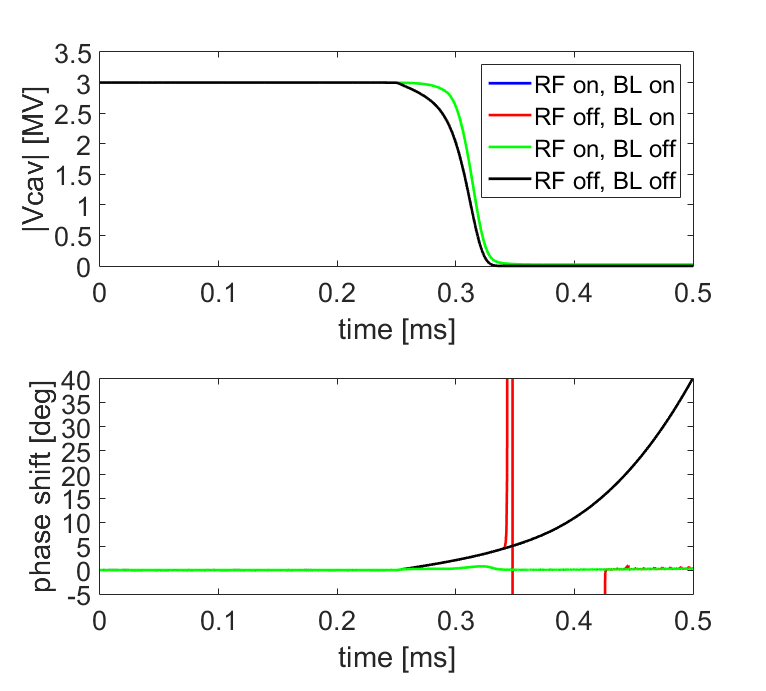}
   \caption{Plots of cavity voltage amplitude (top) and phase (bottom) for an HL-LHC cavity at 2K during a quench, comparing the effects of the RF system and beam loading.}
   \label{fig:paramcav5}
\end{figure}

The same study was repeated for the HL-LHC cavity, using the parameters in Table~\ref{tab:ccparams}. Figures~\ref{fig:paramcav4} and \ref{fig:paramcav5} show the cavity voltage and phase when cooled in 4K and 2K LHe respectively, comparing the effects of beam loading and keeping the RF on or off during a quench.

\begin{figure}
\centering
   \includegraphics*[width=82.5mm]{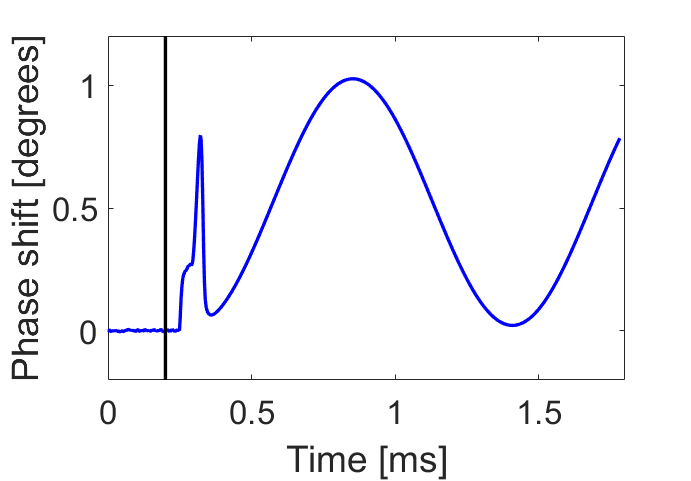}
   \caption{Cavity phase vs. time during a quench for an HL-LHC crab cavity under normal operating conditions.}
   \label{fig:paramcav6}
\end{figure}

We note once again that beam loading has very little effect on the cavity voltage except for causing jumps in the cavity phase when the RF is switched off and only after the quench. Therefore for all further simulations, beam loading will be neglected. Figure~\ref{fig:paramcav6} shows the cavity phase for the HL-LHC crab cavity during a quench under normal operating conditions. In the absence of microphonics, the phase changes by $<1^{\circ}$ with the RF kept on and the LHe temperature at 2K.

\begin{figure}
\centering
   \includegraphics*[width=82.5mm]{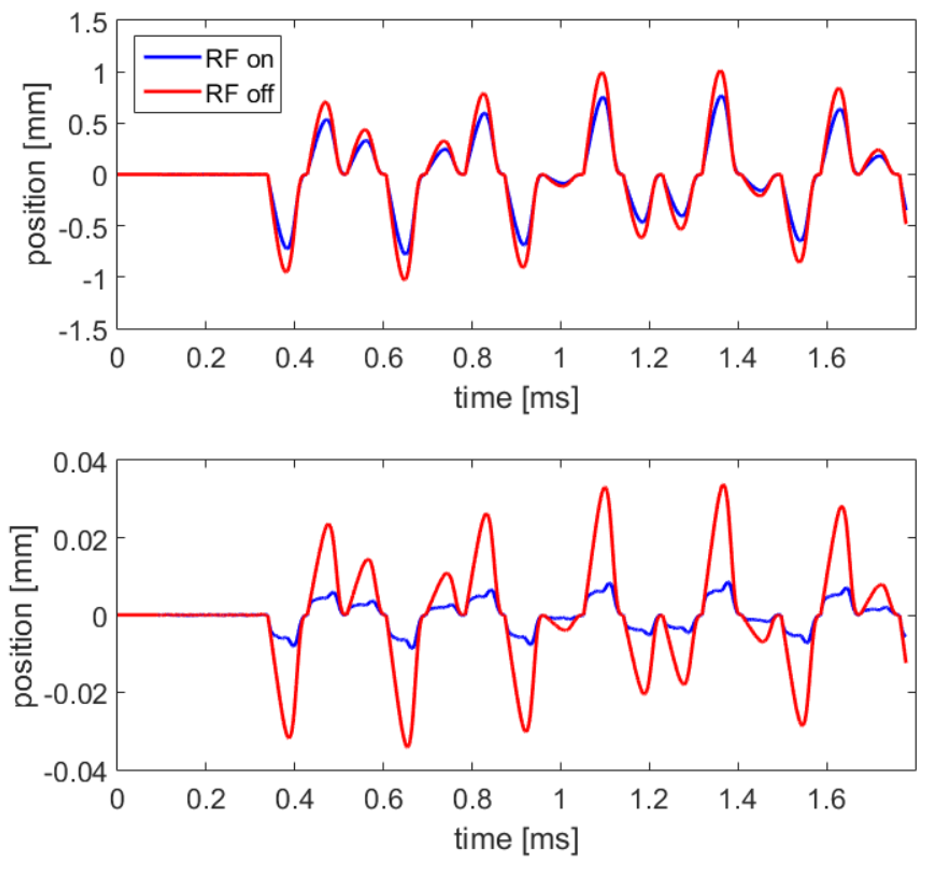}
   \caption{Transverse beam position during a quench in 4K LHe (top) and 2K LHe (bottom) for an HL-LHC cavity, with the RF kept on (blue) or switched off (red) when a quench is detected.}
   \label{fig:paramcav7}
\end{figure}

Figure~\ref{fig:paramcav7} shows the transverse beam position during a quench when the LHe temperature is 4K (top) and 2K (bottom). Due to the much higher energy of the HL-LHC beam compared to KEKB, the beam, hence reference particle, experiences significantly less deflection.

\subsection{SixTrack simulation results}
\label{sec:results:sixtrack}
For both the SPS and HL-LHC, 3 different cases were studied with SixTrack.
They will hereafter be named as follows:
\begin{description}
\item[Const] The crab cavity voltage amplitude and RF phase at the time of arrival for the ideal particle are constant throughout the simulation. This is meant as a reference of comparison for the other losses
\item[NoFail] The crab cavity voltage and phase are constant for 20 (SPS) or 100 (HL-LHC) turns in the beginning of the simulation in order to make sure that the beam has stabilized, and then follow a set of pre-calculated values corresponding to a cavity experiencing microphonics but no quench, being controlled by an LLRF controller for the remaining 100~or 20~turns.
The modelled cavity is a LHC-type cavity at 2K, as described in Table~\ref{tab:ccparams}.
\item[Fail] The crab cavity voltage and phase are constant in the beginning of the simulation, and later follow a set of pre-calculated values corresponding to a cavity experiencing microphonics and then quenching with presure detuning.
The quench starts at 200~{\textmu}s after the ``unlocking'' of the crab cavity, and the pressure detuning occurs after 250~{\textmu}s.
The modelled cavity is a LHC-type cavity at 2K, as described in Table~\ref{tab:ccparams}.
\end{description}
In all cases, the initial voltage and phase of the controlled CC was set to 3~MV and 0$^\circ$ respectively.
The time-developement of the RF voltage and phase are shown in Figures~\ref{FIG:SixTrack:LHC:LLRFSIM} (LHC) and~\ref{FIG:SixTrack:SPS:LLRFSIM} (SPS), for both the NoFail and Fail scenarios. These plots start from the time where the controlled cavity is ``unlocked'', which is considered as $t=0$ and the beginning of turn 1.
As shown in the plots, in the NoFail scenario the voltage remains constant at $3~\mathrm{MV} \pm 3~\mathrm{kV}$, while the swings within $\pm 2^\circ$ due to microphonics, both for the SPS and LHC cases; for the Fail scenario we see the same behaviour up to 200~{\textmu}s when the quench happens, the voltage then stays approximately stable for 80~{\textmu}s until it drops to zero within 60~{\textmu}s, while the phase has an excursion up to 6.7$^\circ$ due to a combination of the pressure and Lorentz detuning, and microphonics. 


\subsubsection{HL-LHC machine}
\label{sec:results:sixtrack:LHC}

\begin{figure*}
  \centering
  \includegraphics*[width=85mm]{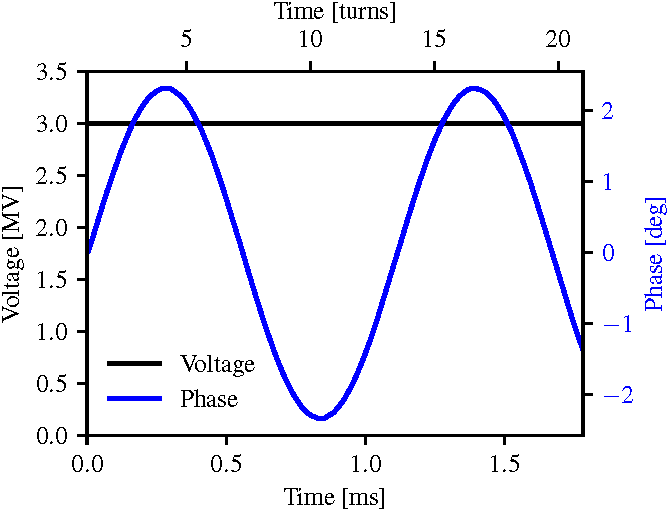}
  \hfill
  \includegraphics*[width=85mm]{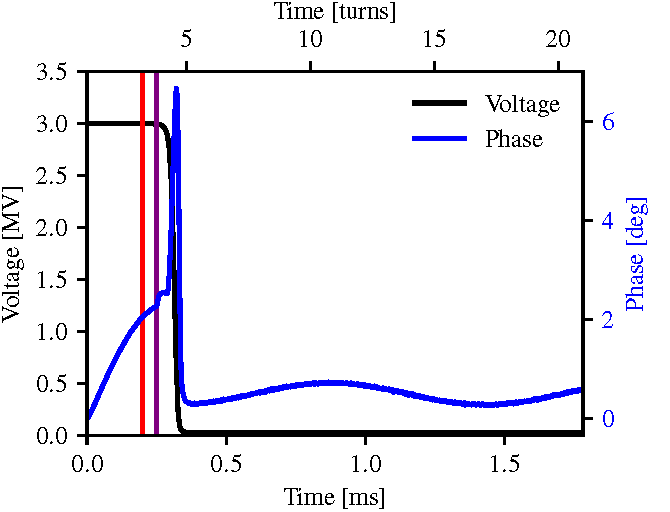}
  \caption{Crab cavity voltage and phase as a function of time for the HL-LHC, both NoFail (left) and Fail(right) as defined in Section~\ref{sec:results:sixtrack}. For the Fail scenario, the first vertical line (red) marks the beginning of the quench, and the second line (purple) the time when pressure detuning becomes important.}
  \label{FIG:SixTrack:LHC:LLRFSIM}

  
  \centering
  \includegraphics[width=85mm]{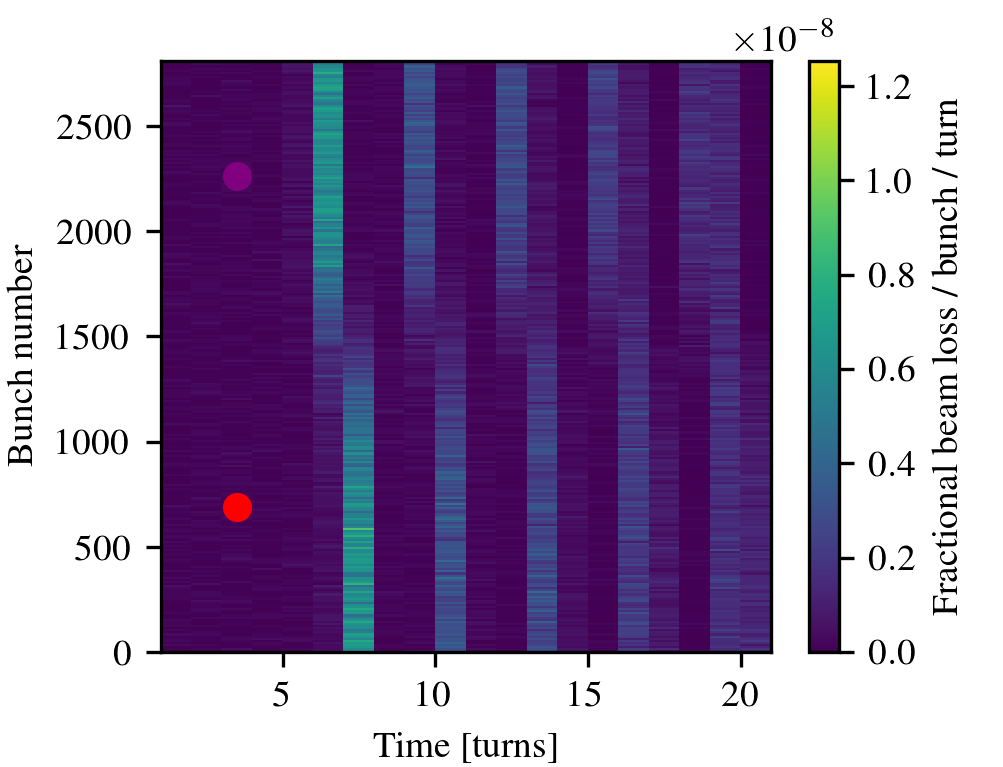}
  \hfill
  \includegraphics[width=85mm]{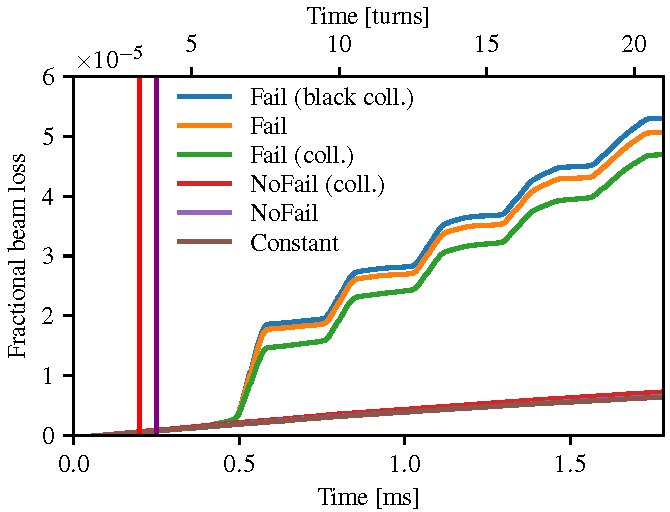}
  \caption{Losses in the HL-LHC as a function of time; both instantaneous losses (left) for the Fail scenario, and cumulative losses comparing different scenarios and simulation techniques (right). The first vertical circle or line (red) marks the beginning of the quench, and the second circle or line (purple) the time when pressure detuning becomes important for the Fail scenario.}
  \label{FIG:SixTrack:LHC:LossesTime}
  %
  %
  %

\end{figure*}

\begin{figure*}
  \centering
  \includegraphics[width=85mm]{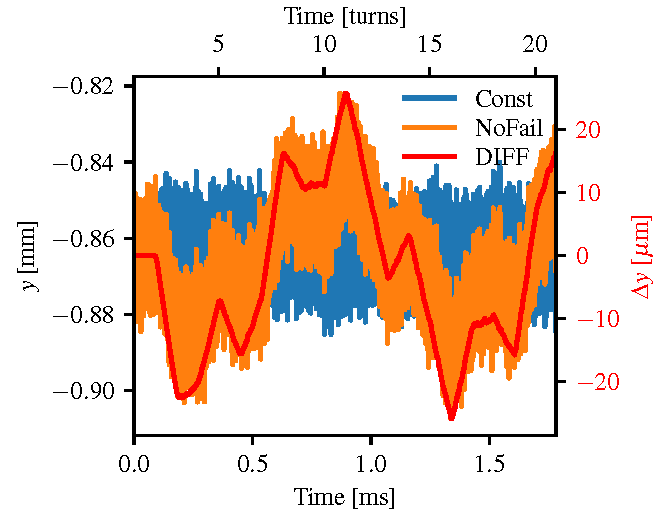}
  \hfill
  \includegraphics[width=85mm]{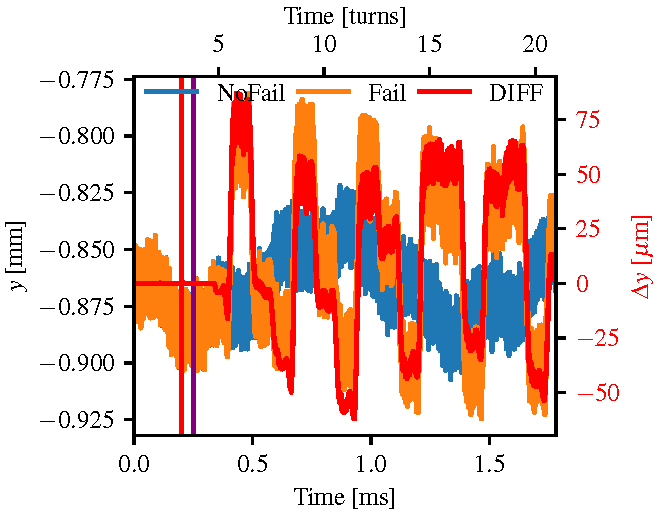}

  \includegraphics[width=85mm]{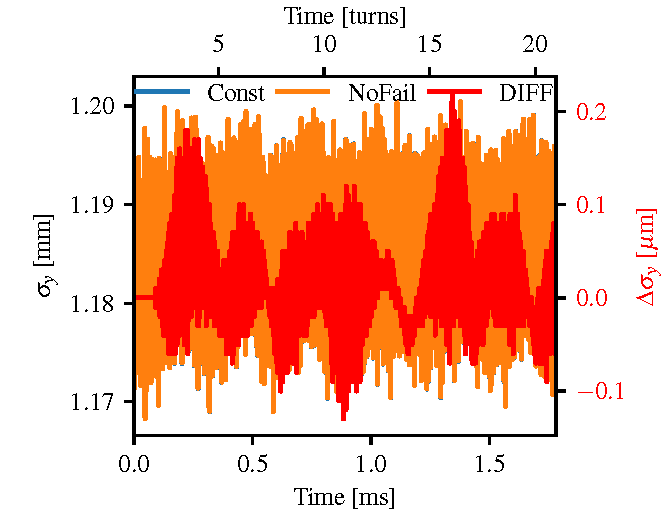}
  \hfill
  \includegraphics[width=85mm]{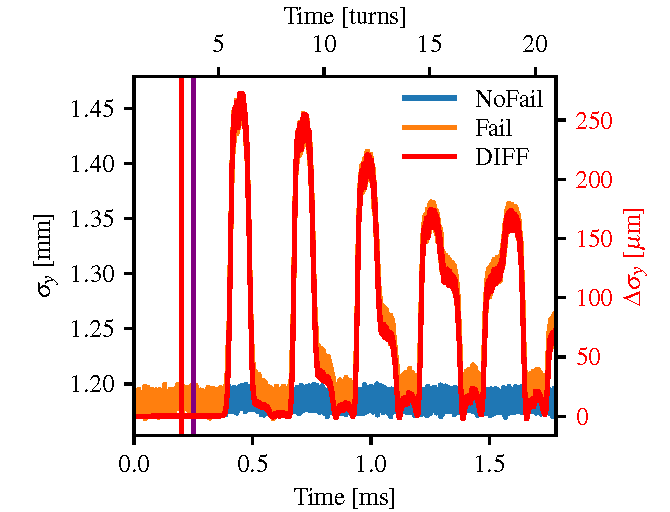}
  \caption{Vertical beam position (top) and size (bottom) at the controlled crab cavity in the LHC, comparing scenarios NoFail vs.\ Const (left) and Fail vs.\ NoFail (right). For the Fail scenario, the first vertical line (red) marks the beginning of the quench, and the second line (purple) the time when pressure detuning becomes important.}
  \label{FIG:SixTrack:LHC:V}

\end{figure*}

For the LHC, the cavity parameters were always constant for the first 100~turns, and then controlled by the LLRF system model for the last 20~turns.
The voltage and phase during these last 20 turns of the Fail- and NoFail cases are shown in Figure~\ref{FIG:SixTrack:LHC:LLRFSIM}.

HL-LHC version~1.2 optics and sequence for beam~1 was used~\cite{BejarAlonso:2069130,Apollinari:2116337}, which includes 4 cavities per beam per IP per side.
Only one of the crab cavities, named ``ACFCA.AL1.B1'', which is the cavity on beam~1 ``upstream'' of IP1 that is closest to the IP, was controlled by the LLRF system and experienced the quench.
All other cavities were assumed to have a constant voltage and phase; the other upstream cavities have a voltage of 3.0~MV, while for the downstream cavities the voltage is matched using MadX~\cite{madx} to 2.871~MV for IP1 and 2.902~MV for IP5, in order to minimize orbit error through the ring during normal operation.

As mentioned previously, both collimation SixTrack and the standard SixTrack with an external aperture check was used to simulate the Fail and NoFail scenarios. This allowed comparing the losses between the two cases, checking that the specially developed external aperture check was functioning correctly, as the use of this code was a neccessity for the SPS simulations.
Using the external aperture check, the particle population was trimmed at the three primary collimators ``TCP.D6l7.b1'', ``TCP.C6l7.b1'', and ``TCP.B6l7.b1'', which are located in IP7 and have an aperture of 5.7~$\sigma$ relative to a normalized emittance of 3.5~{\textmu}m; equivalently, the collimator aperture is 6.74~$\sigma$ relative to the actual normalized beam emittance of 2.5~{\textmu}m.
As seen in Figure~\ref{FIG:SixTrack:LHC:LossesTime} (right), the two methods are quite comparable, with the external aperture check predicting slightly larger losses than the collimation version.
This is expected as the external check does not take particle scattering into account, so that all particles touching the collimator are lost.
Furthermore, if the material of the primary collimators in the collimation simulation is changed from carbon to a perfect black absorber, the predicted losses are larger than those predicted by the external aperture check.
This is also expected since the collimation simulation takes the length of the collimators into account, while the external aperture check only checks the losses at a sigle plane.
Finally, the losses are found to be on the order of $3\cdot10^{-5}$ of the full beam 10 turns after the failure, which is comparable to what has been seen in similar simulations previously~\cite{Andreaphdthesis,TalkKyrreParis}.
This level of losses would be survivable for the collimation system~\cite{Andrea15,Bruning:782076}; however this is greatly influenced by the betatron tune and the phase advance between the failing crab cavity and the primary collimators~\cite{Andreaphdthesis}.

As seen in the upper plots of Figure~\ref{FIG:SixTrack:LHC:V}, the cavity failure creates a movement of up to  $\approx\pm~50$~{\textmu}m in the vertical plane at the point of the cavity. As expected, the frequency of the beam position oscillation is the vertical betatron tune; and as seen in Figure~\ref{FIG:SixTrack:LHC:LossesTime} (left), this frequency is also observed in the time profile of the losses, where the beam touches a collimator jaw every 3 turns.
The centroid motion is caused by the phase error while the cavity voltage has not yet dropped to zero; however as this phase error is on the order of a few degrees, we observe a relatively small motion.
In the horizontal plane, the centroid motion is only $\approx\pm2$~{\textmu}m; this motion is caused by coupling from the vertical plane.

For the non-failed case there is still a movement up to $\approx\pm~20$~{\textmu}m due to the phase variation. 

For the vertical beam size, shown in Figure~\ref{FIG:SixTrack:LHC:V} (bottom), we see that it increases from approximately 1.2~mm to 1.45~mm, an increase of 20\%.
As discussed in more detail in the analysis of the SPS data, this is caused by different $z$-slices of the bunch seeing a different kick due to the non-cancellation between the cavities creating and closing the bump, and are thus traveling on different trajectories.
This results in the projection of the bunch onto the $y$-axis being wider, and thus a larger vertical beam size.
The horizontal beam size remains constant within 1~{\textmu}m in all cases.

\subsubsection{SPS machine}
\label{sec:results:sixtrack:SPS}

\begin{figure*}[htb!]
  \centering
  \includegraphics*[width=85mm]{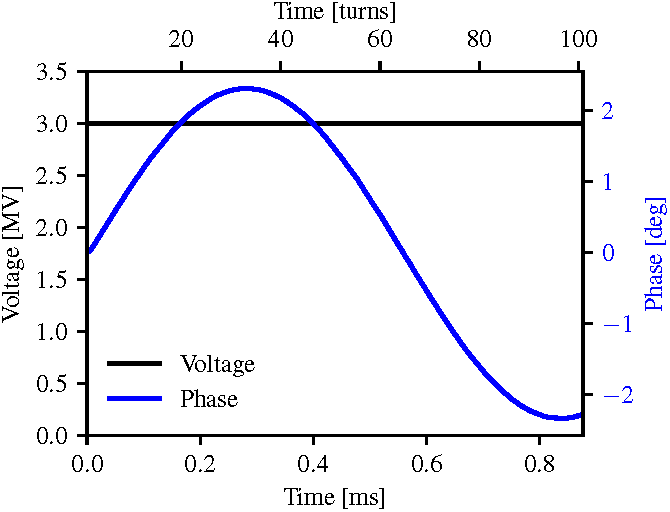}
  \hfill
  \includegraphics*[width=85mm]{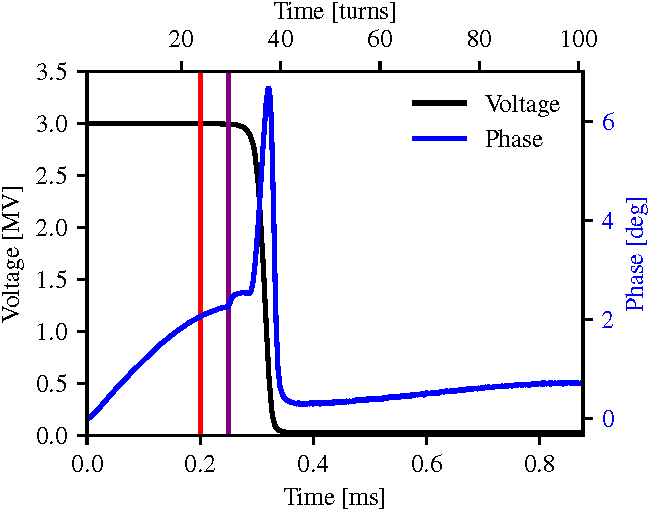}
  \caption{Crab cavity voltage and phase as a function of time for the SPS, both NoFail (left) and Fail (right) as defined in Section~\ref{sec:results:sixtrack}. For the Fail scenario, the first vertical line (red) marks the beginning of the quench, and the second line (purple) the time when pressure detuning becomes important.}
  \label{FIG:SixTrack:SPS:LLRFSIM}
\end{figure*}
  
\begin{figure*}
  \centering
  \includegraphics[width=85mm]{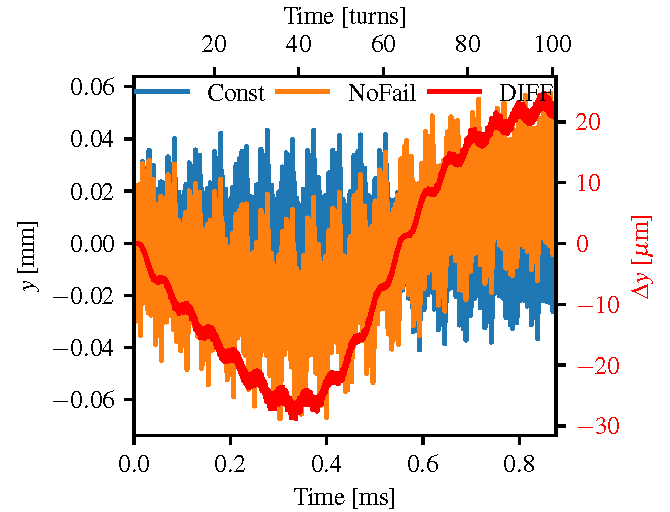}
  \hfill
  \includegraphics[width=85mm]{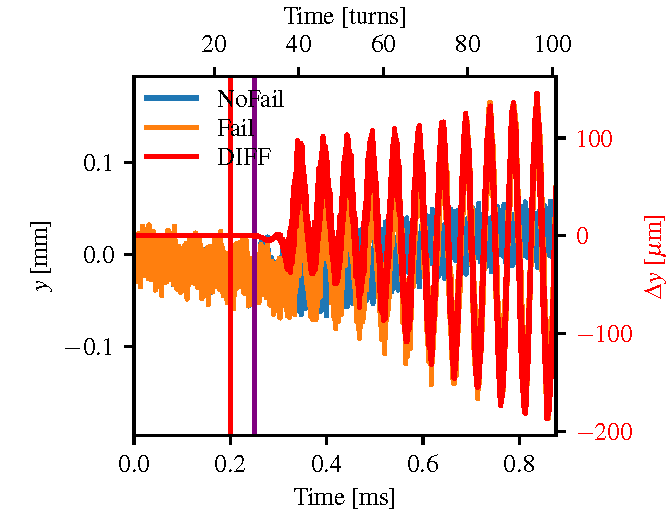}
  
  \includegraphics[width=85mm]{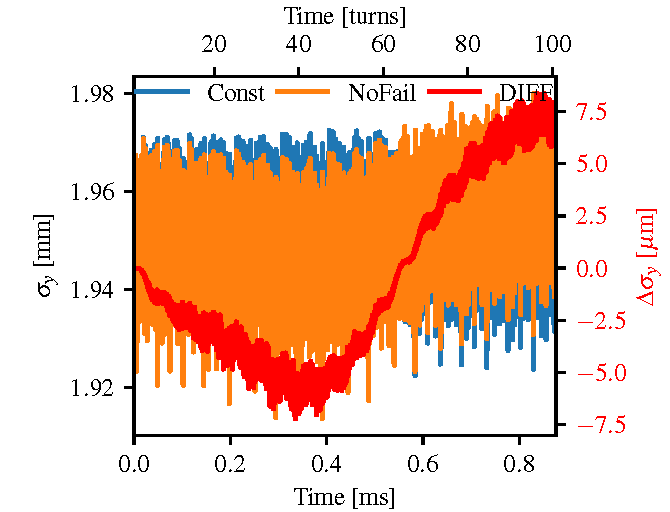}
  \hfill
  \includegraphics[width=85mm]{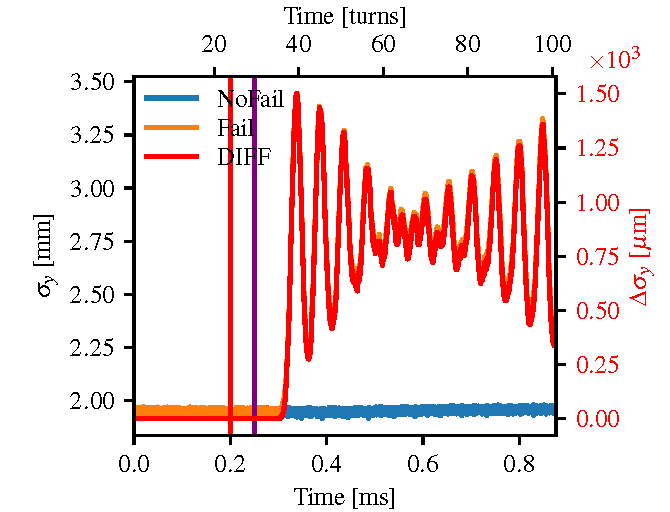}
  
  \caption{Vertical beam position (top) and size (bottom) at the controlled crab cavity in the SPS / 55~GeV, comparing NoFail vs.\ Const (left) and Fail vs.\ NoFail (right). For the Fail scenario, the first vertical line (red) marks the beginning of the quench, and the second line (purple) the time when pressure detuning becomes important.}
  \label{FIG:SixTrack:SPS:V}
\end{figure*}

The crab cavities have never been used in proton machines and is therefore crucial to ensure that there will be no detrimental side effect for the beam in the HL-LHC machine.
With this in mind, a set of prototype vertical crab cavities have been installed first in the SPS, and served as a test-bed between May to November 2018. 

Prior to the crab cavity installation, dedicated Machine Development (MD) studies were undertaken in order to ensure good understanding of the SPS machine and the limitation of its instruments.
Due to the limited available time for MDs, good preparation and planning of the studies was essential. 

The quenching of the SPS crab cavity was studied using the nominal (Q26) SPS optics, the main parameters of which are summarised in Table~\ref{TAB:SixTrack:MACHINEPARAMETERS}.
The emittance was chosen based on the initial emittance at 55~GeV~\cite{Calaga:2012zz}.
Two different beam energies were studied, 55~GeV and 120~GeV, and for each energy, 3 simulations were performed as discussed above.
The controlled cavity was the first crab cavity along the ring, and the second one, installed 1.05~m later, was held a constant voltage of 3.0~MV and constant phase of $0^\circ$.

In order to study the losses in the SPS, the two aperture bottlenecks, the momentum scraper ``TIDP'' and a beam dump ``TIDV'', were included as described above.
The \{H,V\} apertures of these bottlenecks are \{41,15\}~mm for the momentum scraper, which corresponds to a relative aperture of \{19.8,13.06\}~$\sigma$; and \{42.5,20.4\}~mm for the beam dump, which corresponds to \{34.77,10.48\}~$\sigma$.
The listed relative aperture values are calculated for the 55~GeV case; for the 120~GeV case the geometrical emittance is smaller and thus the relative aperture is even larger.
No losses were observed from either the core or the halo, in either aperture bottlenecks.

From Figure~\ref{FIG:SixTrack:SPS:V} (top left), we see that in the NoFail case (orange line) there is a small oscillation, which is due to the phase error caused by the microphonics which causes the orbit to change adiabatically; the period is approximately the same as the phase oscillation of cavity.
Comparing the Fail and NoFail (top right plot), we can see that soon after the point where the cavity quenches, the amplitude of the vertical oscillation increases rapidly.
Note that even for this case there are no losses observed in the SPS simulation, as the apertures bottlenecks are very large.

Unlike the vertical case, the horizontal centroid is constant in time in both Const and NoFail cases.
In the Fail case, some very small oscillation is observed also in the horizontal centroid; this is due to the coupling between the vertical and horizontal planes that is introduced by the sextupoles installed in the SPS lattice.

\begin{figure*}
  \centering
  \includegraphics[width=85mm]{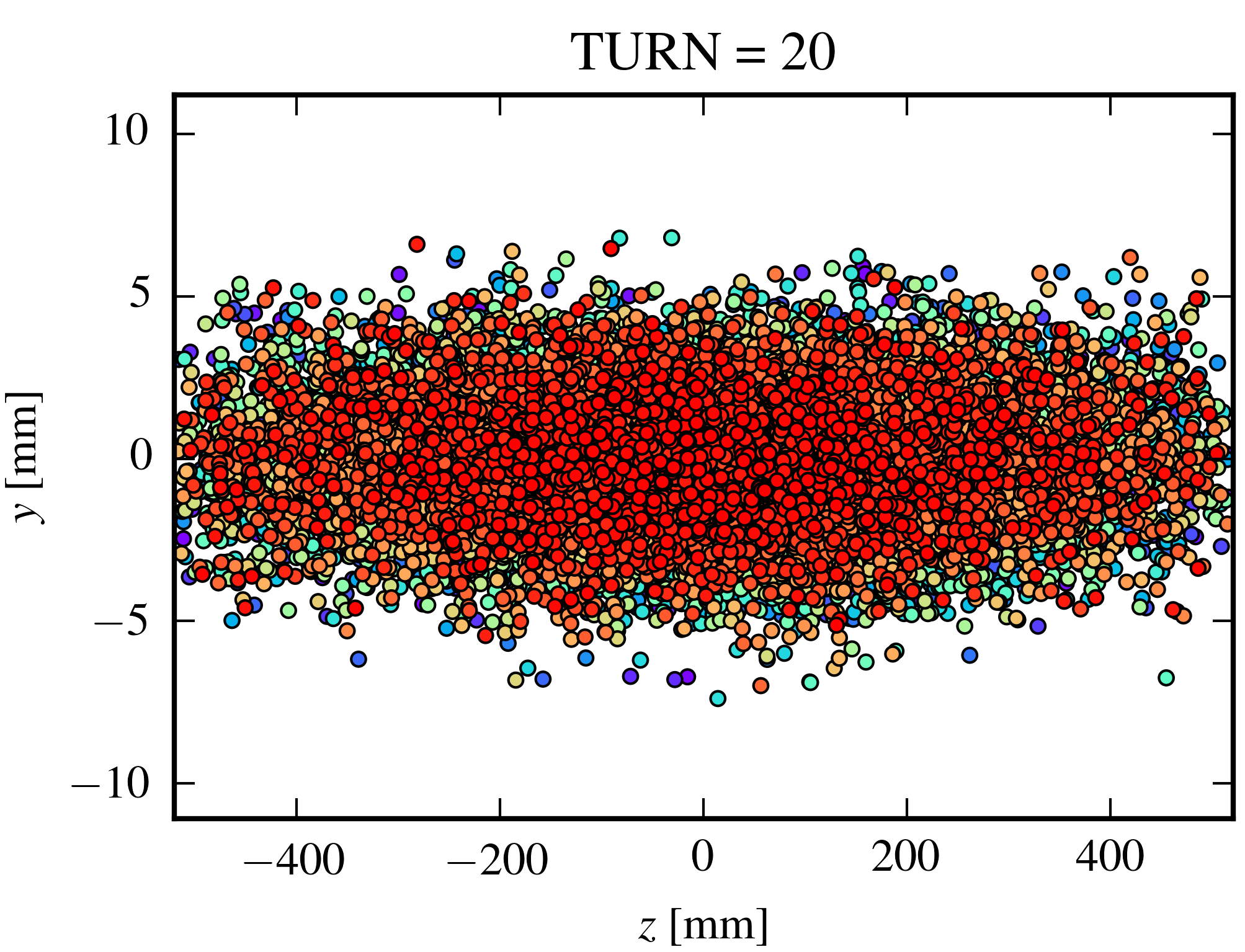}
  \hfill
  \includegraphics[width=85mm]{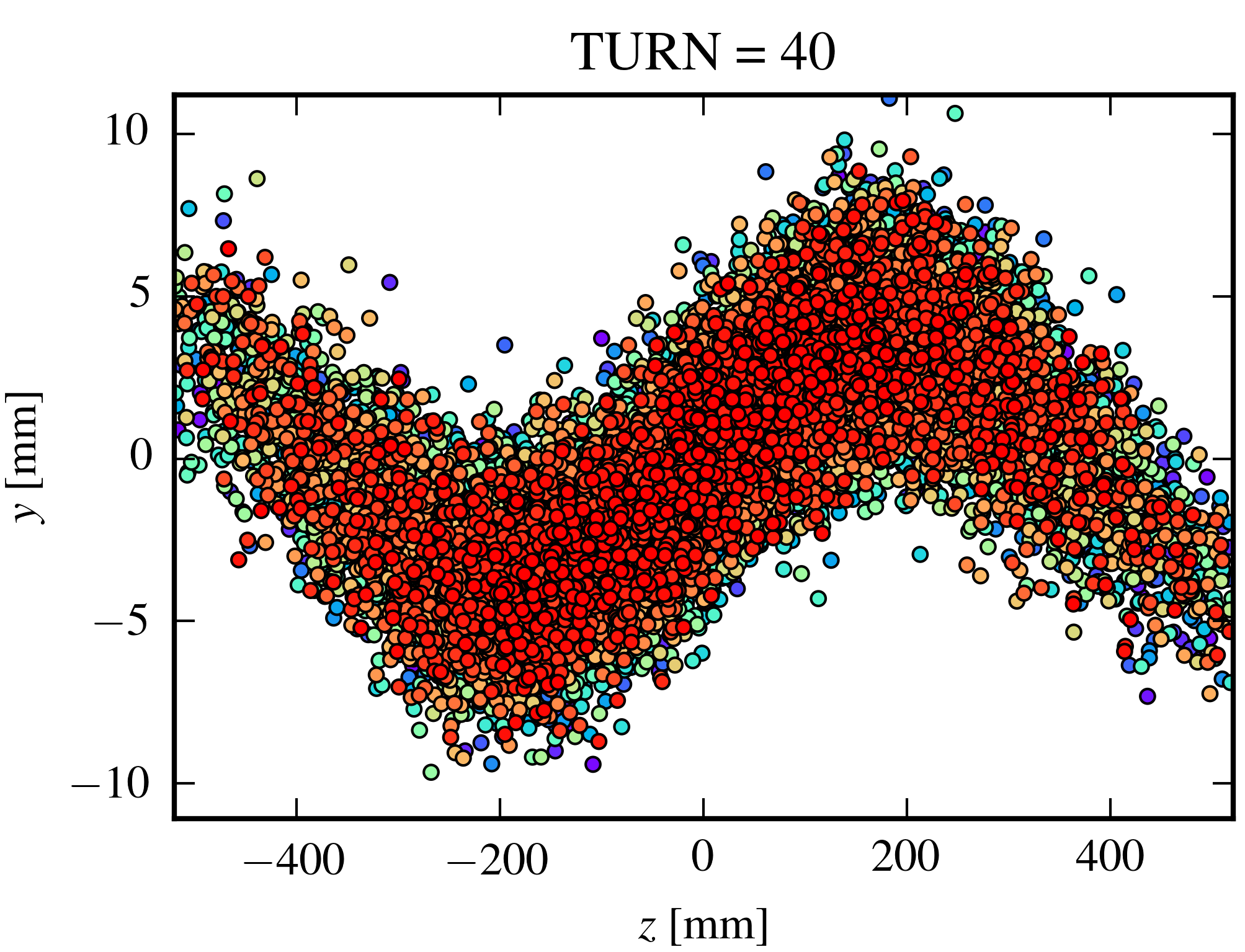}
  
  \includegraphics[width=85mm]{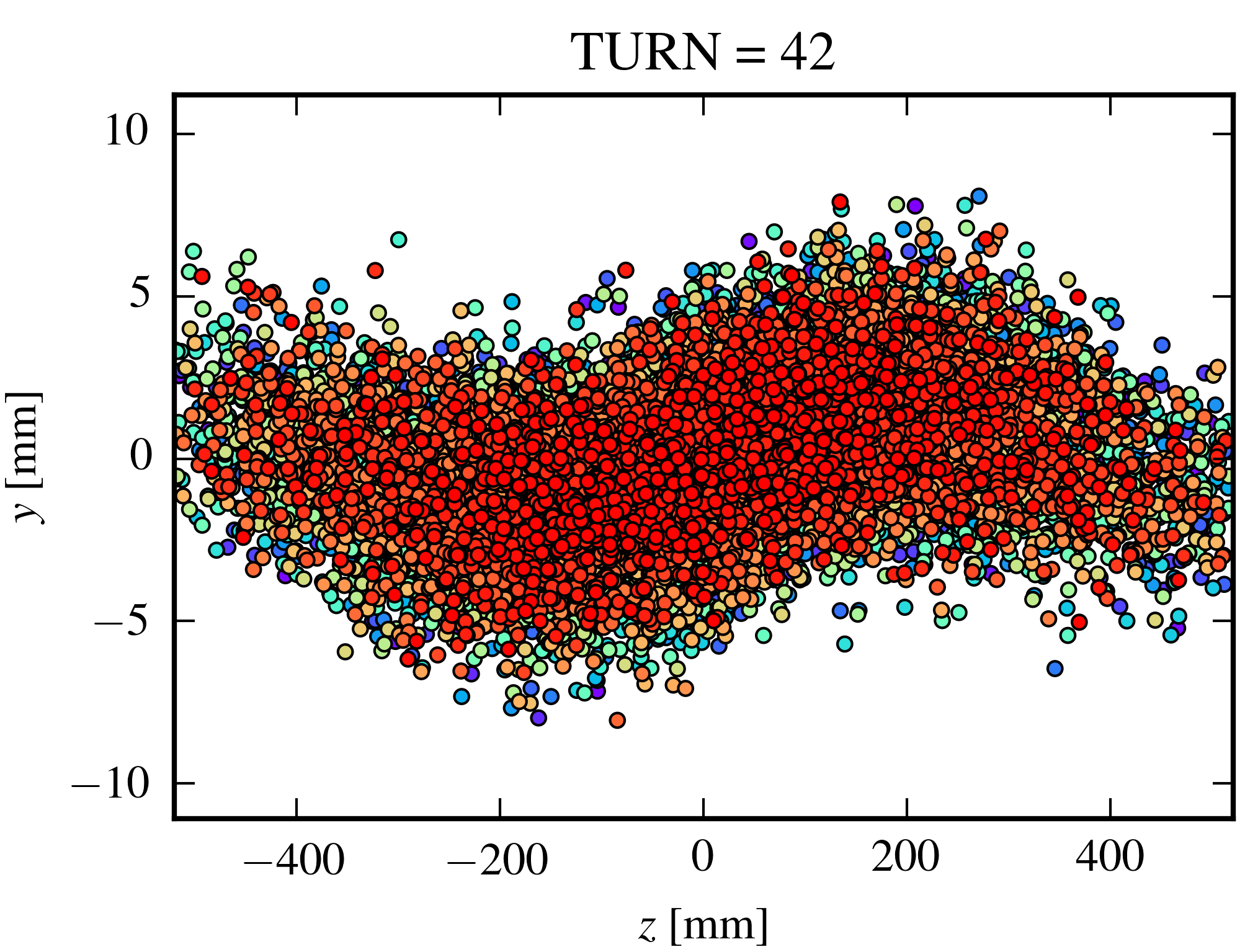}
  \hfill
  \includegraphics[width=85mm]{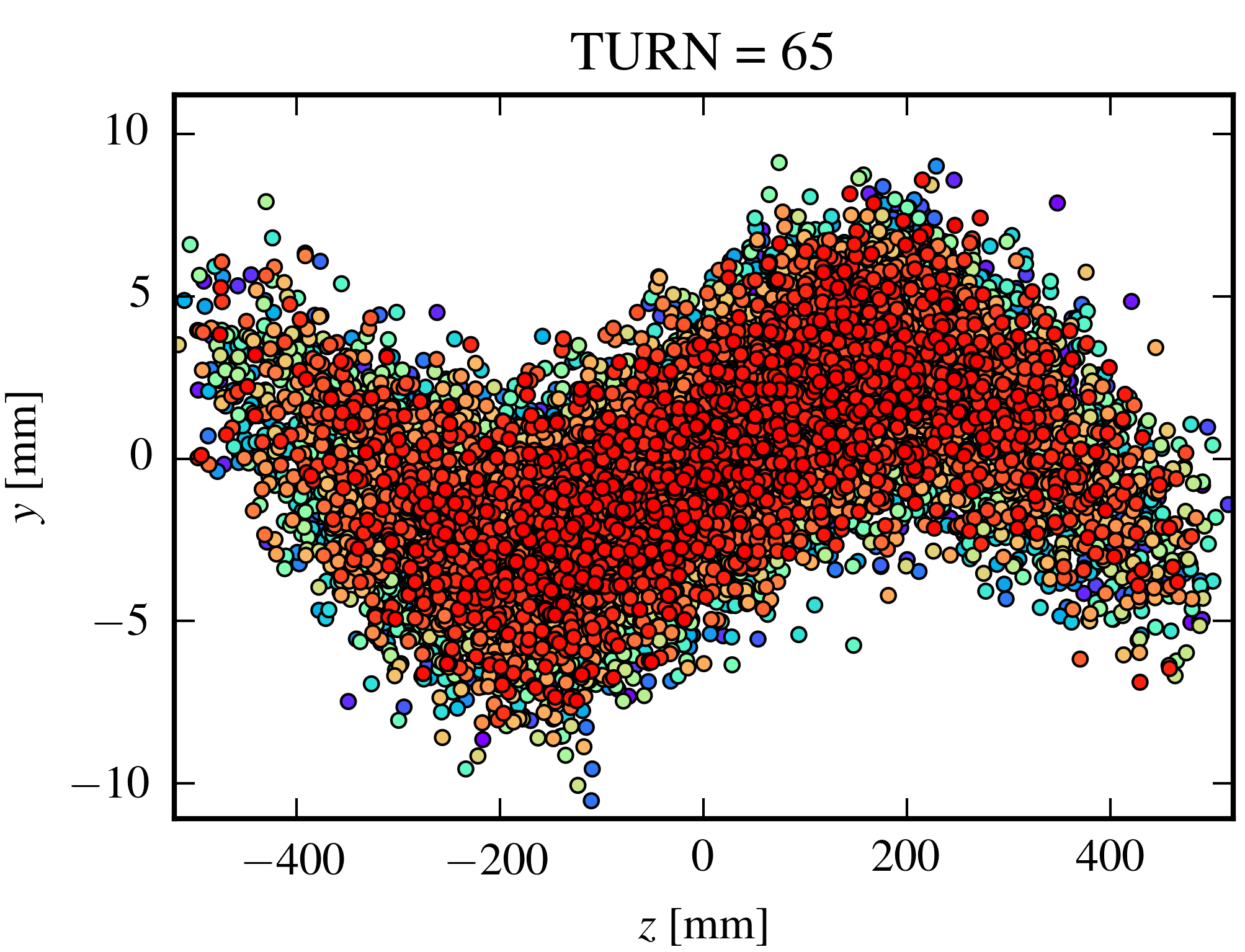}
  
  \caption{Projection of the particle population of bunch 1 into the $\{z,y\}$ plane at 4 characteristic points in time, for the SPS at 55~GeV failure case at the point of the crab cavity under study. Upper left: Turn 20, before the quench; Upper right: Turn 40, near the top of the first peak seen in the left-hand plots of Figure~\ref{FIG:SixTrack:SPS:V}; Lower left: Turn 42, near the bottom of the first minimum just after that peak; Lower right: Turn 65, near the stable region seen around 0.6~ms, where the $y$ projection is quite stable between different turns.}
  \label{FIG:SixTrack:SPS55:VZprojection}
\end{figure*}

There is a clear increase in the vertical beam size when the cavity quenches, as shown in Figure~\ref{FIG:SixTrack:SPS:V} (bottom right); while for the NoFail there is little effect of the phase oscillation (bottom left plot).
The oscillation in the Fail case is due to the head and tail of the bunch having a similar $y$ position.
This is confirmed and explained by the evolution of the bunch shape shown in Figure~\ref{FIG:SixTrack:SPS55:VZprojection}.

The frequency and amplitude modulation of the oscillations of the vertical beam size observed in Figure~\ref{FIG:SixTrack:SPS:V} (bottom right) after the quench is due to the $z$-dependence of the closed orbit, which for a given particle moves with the synchrotron oscillation.
Particles that before the quench were at the head or tail of the bunch, after the quench found themselves far away from their closed orbit, and were thus effectively injected into a large amplitude betatron oscillation around this new closed orbit.
When the head and tail are opposite in phase of their betatron motions, this results in large $\sigma_y$ when they are furthest away from $y=0$ (Figure~\ref{FIG:SixTrack:SPS55:VZprojection}, turn~40), and small $\sigma_y$ when they are both close to $y=0$ (Figure~\ref{FIG:SixTrack:SPS55:VZprojection}, turn~42).
This causes the fast oscillation with frequency equal to the betatron frequency, as seen in Figure~\ref{FIG:SixTrack:SPS:V} (bottom right).
As the particles that were originally (before the quench) at the head and tail drift towards $z=0$ due to the synchrotron oscillation, their closed orbit moves towards $y=y'=0$, adiabatically ``dragging'' the particles along.
This means that their $y$-position is now only caused by the the betatron oscillation, with little addition contribution from the orbit.
At the same time, the particles that were originally close to $z=0$ are also moved adiabatically outwards in $\{y,y'\}$ phase space, resulting in the betatron amplitudes not increasing as much.
This causes the modulation of the amplitude of the oscillation, which is four times the frequency of the synchrotron tune.

For the horizontal beam size no clear effect of the failure is observed, apart from a very small oscillation which is most likely due to the initial distribution not being absolutely perfectly matched.

Similar behaviour is observed between the 55~GeV and 120~GeV cases.
In both cases, the optics and thus the betatron tune is the same, which leads to the oscillations of the centroids staying in phase.
The main difference is, as expected, the amplitude of the oscillations and beam sizes, as they scale linearly with energy.
The other difference is the synchrotron tune, which is slightly faster for the 55~GeV case.
This also affects the oscillation of $\sigma_y$, where the ``minimum'' that is discussed above comes earlier for the 55~GeV case; this is consistent with the explanation of this effect.

\subsubsection{Comparison of the LHC and the SPS}
\label{sec:results:sixtrack:comparison}

The observed centroid motion of the bunch is due to coherent betatron motion; and as expected, the frequency of the centroid motion is equal to the fractional tune.
This was confirmed by a bunch-by-bunch FFT on the data.
Given the difference in the fractional part of the betatron tunes (see Table~\ref{TAB:SixTrack:MACHINEPARAMETERS}), this causes the centroid motion as a function of the turn number to be much faster in the LHC case.

The crab cavity failure in the SPS case is seen more gradually by the beam than for the LHC case, as the number of revolutions during the voltage collapse is larger (the SPS voltage drops to 0 V within 8~turns, as opposed to less than 1~turn in the LHC case, see Figures~\ref{FIG:SixTrack:LHC:LLRFSIM} and~\ref{FIG:SixTrack:SPS:LLRFSIM}).

Note that while the number of turns tracked after the crab cavity failure is much larger in the SPS (100) than for the LHC (20), the difference in revolution time (see Table~\ref{TAB:SixTrack:MACHINEPARAMETERS}) causes the actual time from the start of the quench to the end of the simulation to be much larger in the LHC case (LHC: 1.5~ms, SPS: 0.5~ms)

Comparing the effect of the maximum kick ($V=3.0~\mathrm{MV}$) on the beam, from Table~\ref{TAB:SixTrack:MACHINEPARAMETERS} we see that for the three cases the maximum kick $\Delta y'$ is very different for the different energies and thus beam rigidity.
However the betatron functions in the relevant plane are much higher for the LHC case than for SPS, resulting in a magnified effect on the beam.
Thus when comparing the maximum kick $\Delta y'$ to $1~\sigma$ of angular displacement $\sqrt{\epsilon_g / \beta_\mathrm{twiss}}$, the resulting normalized kicks are very comparable between the 3 cases.
However since in the SPS case the aperture is much larger, we only see losses in the HL-LHC case.

\section{Conclusions and summary}

In this article we present a model of an RF cavity and include models of beam loading, the low-level RF system, detuning mechanisms and superconducting quenches. These results were bench-marked against experimental measurements of the KEKB crab cavities. After successfully reproducing the key features from the KEKB crab cavity studies, parameter studies were undertaken to identify the causes of the rapid phase shifts observed at KEKB. The same parameter study was also undertaken for the HL-LHC crab cavity to model how it would behave during a quench under different conditions; this is essential for designing suitable machine protection for the crab cavity system.

Full tracking studies with SixTrack were undertaken to obtain a quantitative understanding of the expected beam losses in HL-LHC and SPS due to a quench of one of the crab cavities as well as the evolution of the beam parameters.

Further studies are envisaged to use the cavity model described to study quenches of the HL-LHC accelerating cavities and has been integrated into SixTrack as a module for other applications. This now allows SixTrack to communicate in a general way with external codes and files.




\end{document}